\documentclass[preprint,3p,11pt]{elsarticle}

\usepackage{graphicx,subfigure}
\usepackage{amssymb}
\usepackage{amsmath,mathrsfs}
\usepackage{bm}
\usepackage{url}
\usepackage{xcolor}
\usepackage{lipsum}

\journal{Journal of Computational Physics} 

\numberwithin{equation}{section}

\newcommand{\PP}[2]{\frac{\partial#1}{\partial#2}}
\newcommand{\DD}[2]{\frac{\text{d}#1}{\text{d}#2}}
\newcommand{\ii}{\text{i}}
\newcommand{\sinhc}{\,\text{sinhc}}
\newcommand{\sinc}{\,\text{sinc}}

\begin{document}

\begin{frontmatter}

\title{Accurately simulating nine-dimensional phase space of relativistic particles in strong fields}

\author[UCLAEE]{Fei Li\corref{cor}}
\ead{lifei11@ucla.edu}
\cortext[cor]{Corresponding author}
\author[UCLAPH]{Viktor K. Decyk}
\author[UCLAPH]{Kyle G. Miller}
\author[UCLAPH]{Adam Tableman}
\author[UCLAPH]{Frank S. Tsung}
\author[IST]{Marija Vranic}
\author[IST,ISCTE]{Ricardo A. Fonseca}
\author[UCLAEE,UCLAPH]{Warren B. Mori\corref{cor}}
\ead{mori@physics.ucla.edu}

\address[UCLAEE]{Department of Electrical Engineering, University of California Los Angeles, Los Angeles, CA 90095, USA}
\address[UCLAPH]{Department of Physics and Astronomy, University of California Los Angeles, Los Angeles, CA 90095, USA}
\address[IST]{GOLP/Instituto de Plasma e Fus\~ao Nuclear, Instituto Superior T\'ecnico, Universidade de Lisboa, Lisbon, Portugal}
\address[ISCTE]{ISCTE - Instituto Universit\'ario de Lisboa, 1649--026, Lisbon, Portugal}

\begin{abstract}
Next-generation high-power laser systems that can be focused to ultra-high intensities exceeding $10^{23}$~W/cm$^2$ are enabling new physics regimes and applications. The physics of how these lasers interact with matter is highly nonlinear, relativistic, and can involve lowest-order quantum effects. The current tool of choice for modeling these interactions is the particle-in-cell (PIC) method. In the presence of strong electromagnetic fields, the motion of charged particles and their spin is affected by radiation reaction (either the semi-classical or the quantum limit). Standard (PIC) codes usually use Boris or similar operator-splitting methods to advance the particles in standard phase space. These methods have been shown to require very small time steps in the strong-field regime in order to obtain accurate results.
In addition, some problems require tracking the spin of particles, which creates a nine-dimensional (9D) particle phase space, i.e., $(\mathbf{x},\mathbf{u},\mathbf{s})$. Therefore, numerical algorithms that enable high-fidelity modeling of the 9D phase space in the strong-field regime (where both the spin and momentum evolution are affected by radiation reaction) are desired. We present a new particle pusher that works in 9D and 6D phase space (i.e., with and without spin) based on analytical rather than leapfrog solutions to the momentum and spin advance from the Lorentz force, together with the semi-classical form of radiation reaction in the Landau-Lifshitz equation and spin evolution given by the Bargmann-Michel-Telegdi equation. Analytical solutions for the position advance are also obtained, but these are not amenable to the staggering of space and time in standard PIC codes. These analytical solutions are obtained by assuming a locally uniform and constant electromagnetic field during a time step. The solutions provide the 9D phase space advance in terms of a particle's proper time, and a mapping is used to determine the proper time step duration for each particle as a function of the lab frame time step. Due to the analytical integration of particle trajectory and spin orbit, the constraint on the time step needed to resolve trajectories in ultra-high fields can be greatly reduced. The time step required in a PIC code for accurately advancing the fields may provide additional constraints. We present single-particle simulations to show that the proposed particle pusher can greatly improve the accuracy of particle trajectories in 6D or 9D phase space for given laser fields. We have implemented the new pusher into the PIC code \textsc{Osiris}. Example simulations show that the proposed pusher provides improvement for a given time step. A discussion on the numerical efficiency of the proposed pusher is also provided.
\end{abstract}


\end{frontmatter}

%% main text
\section{Introduction}

With the recent advent of petawatt-class lasers and a roadmap for multi-petawatt-class laser systems~\cite{ELI,XCELS,SULF}, laser intensities exceeding $10^{23}$~W/cm$^2$ will soon become available. These lasers will open a new door for research avenues in plasma physics, including plasma-based acceleration~\cite{tajima1979,chen1985,joshi2003,lu2007} in the strong-field regime, the coupling of laser-plasma interactions and quantum electrodynamics (QED)~\cite{vranic2016a},  and the ability to mimic some astrophysical phenomena (e.g., gamma-ray bursts and supernova explosions) in the laboratory. The physics of how ultra-high-intensity lasers interact with matter is highly nonlinear, relativistic, and involves non-classical processes such as radiation reaction and quantum effects.  Simulations will be a critical partner with experiments to unravel this physics. The electromagnetic particle-in-cell (PIC) algorithm~\cite{dawson1983,hockney1988,birdsall2005} has been successfully applied to the research of plasma or charged-particle beams interacting with radiation for nearly half a century. With moderate radiation (laser) parameters, e.g., $eA/(m_ec^2)\gtrsim1$, where $A$ is the vector potential of the laser, PIC simulations have proven to be a reliable tool. However, in the strong-field regime where $eA/(m_ec^2)\gg1$, accurate modeling becomes much more challenging. Developing high-fidelity PIC simulation algorithms requires a comprehensive and deep understanding of each aspect of the numerical algorithm and the physical problem itself. To improve the simulation accuracy and reliability, much effort has already been undertaken to mitigate various numerical errors; these include improper numerical dispersion, errors to the Lorentz force for a relativistic particle interacting with a laser,
numerical Cerenkov radiation and an associated instability~\cite{godfrey2013,xu2013,yu2015a,yu2015b,li2017}, finite-grid instability~\cite{langdon1970,okuda1972,meyers2015,huang2016} and spurious fields surrounding relativistic particles~\cite{xu2020}.

In this article, we address inaccuracies and challenges for the particle pusher used as part of a PIC code. The pusher has been found to be one of the major factors that prevent high-fidelity PIC simulations in the strong-field regime. Most PIC codes use the standard Boris scheme~\cite{boris1972} or one of its variants~\cite{vay2008,higuera2017} for the particle push. These later variants correct a shortcoming of the Boris push for a particle moving relativistically where $\mathbf E + \mathbf v \times \mathbf B \approx 0$. In the standard Boris split algorithm, the velocity can change even when the Lorentz force vanishes. However, when the fields (forces) are large, these algorithms require small time steps to provide sufficient accuracy. 

Gordon et al.~\cite{gordon2017} showed that it is possible to construct an analytic or exact covariant non-splitting pusher. This method assumes the fields (forces) are constant during an interval of the proper time and then advances the particle momentum using analytic solutions. Since this method pushes particles in the proper time rather than in the observer's time, it cannot be directly applied to PIC simulations and can only be used for single-particle tracking. Gordon et al.  also discussed how to include radiation reaction (RR), but used a form of RR that is challenging to incorporate. In some very recent  work~\cite{gordon2021}, Gordon and Hafizi propose a more compact in form which they call a special unitary particle pusher. This method provides a method to obtain solutions to all orders of the time step that maintains Lorentz invariance. They show that with a second-order-accurate mapping from the simulation time step to the proper time step, this pusher can be comparable to the standard Boris pusher in the push rate.
P\'etri~\cite{petri2020} (who does not seem to be aware of the earlier work of Gordon et al.) recently proposed a different implementation of the exact pusher that relies on Lorentz transforming into the particles rest frame and that includes a mapping between the proper and observer time step, allowing the pusher to be applicable for PIC simulations.  However, P\'etri did not consider RR in his implementation.

In strong fields, the motion of charged particles will be significantly impacted by the RR force and its accompanied energy loss. Therefore, determining how to accurately model the RR  effect is also of crucial importance when in the strong-field regime. The Lorentz-Abraham-Dirac (LAD) equation describes the radiation reaction in the semi-classical perspective~\cite{jackson2007}. However, this equation has unphysical runaway solutions that can be avoided by instead using the Landau-Lifshitz (LL) equation~\cite{landau2013}, which was shown to contain all physical solutions of the LAD equation~\cite{spohn2000}. There are other models appropriate for numerical implementation (their comparison can be found in~\cite{vranic2016b}), and most of them give similar results when applied to the semi-classical interaction configurations accessible with near-term laser technology. However, only the LAD and LL models were shown to be consistent classical limits of the QED description of an electron interacting with a strong plane wave~\cite{ilderton2013}.

The numerical integration of electron motion in strong fields requires a very fine temporal resolution, especially when the electron is not ultra-relativistic~\cite{vranic2016b}. This is independent of the choice of the radiation reaction model and is true even when the radiation reaction is turned off. This calls for solutions like sub-cycling~\cite{arefiev2015} or the exact pusher proposed in this manuscript. 
The usual way to implement the additional RR force in PIC codes is (1)~to integrate the particle trajectory using a pusher (splitting or exact) solely for the Lorentz force and then (2)~to add an impulse from the RR force separately~\cite{tamburini2010,gordon2017}. This splitting process is simple to implement but can lead to the accumulation of errors in simulations with a large number of time steps, even though the RR effect is perturbative.
During the review process, one of the reviewers brought to our attention a relatively recent purely theoretical work  by Yaremko \cite{yaremko2013} where  analytical solutions for the four momentum and position vectors are also obtained to the reduced LL equation in the presence of constant electromagnetic fields. The formulae for the particle momentum could be applied in PIC simulations since the constant field is a natural assumption for every time step therein. However, the formulae for the particle position and the mapping between the proper time and the observer time are not amenable for numerical implementation due to their complicated forms. In addition, the singularity of these solutions for appropriate limits needs to be carefully treated, which is critical for numerical implementation.
In this article, we will explore all these elements with an eye toward developing a new particle pusher. We also consider how to couple the evolution of the momentum and spin with RR included. Based on the use of analytical solutions, the proposed pusher is free of numerical errors caused by splitting the operator for the Lorentz force.
Although the motivation for developing an analytic pusher was to handle ultra-high fields, such a pusher will also accurately model the motion of a relativistic particle when $\mathbf E + \mathbf v \times \mathbf B \approx 0$.
The analytic pusher (or any sub-cycling approach~\cite{arefiev2015}) will exhibit some errors in particle trajectories from assuming the fields are constant during a time step. Therefore, the time step must properly resolve the evolution of the fields as well. 
 
The PIC method is also beginning to be used to study the production of spin-polarized particle beams. Furthermore, there is also a growing interest~\cite{li2019,song2019,geng2020} in how RR  affects particle-spin dynamics in strong fields. Particle spin precession follows the Bargmann-Michel-Telegdi (BMT) equation~\cite{jackson2007}, in which the phase space $(\mathbf{x},\mathbf{u})$ is used to evaluate the spin $\mathbf{s}$, where $\mathbf{x}$ and $\mathbf{u}$ are position and momentum, respectively. Therefore, the effect of radiation reaction on the phase space trajectories $(\mathbf{x},\mathbf{u})$ will be also manifested in the behavior of spin dynamics. However, there is far less literature directly related to the numerical schemes of the spin ``push'' than those of the momentum ``push''. A typical numerical method~\cite{vieira2011} is similar to the Boris scheme: the spin orbit is approximated to be a pure rotation with a frequency that is evaluated with the time-centered values of the electromagnetic fields and particle momentum. This Boris-like scheme is subject to large numerical errors in the strong-field regime as will be shown later. In this article, we also derive semi-analytic solutions to the BMT equation by utilizing the analytic expressions of particle momentum without radiation reaction to advance the spin within an interval of time. The RR is then included as an impulse. During the next interval of time, the initial conditions for the analytical update of the momentum are thus different, impacting the spin evolution during subsequent time intervals. Obtaining a fully analytic solution to the BMT equation in the presence of radiation reaction is extremely difficult and likely impossible; however, the RR force can still be accurately included via the aforementioned splitting correction method. We note that this semi-analytic approach can also be applied when quantum effects for RR are included. We leave comparisons of examples with QED for a later publication.

Although we are focusing on finding analytical solutions for both the momentum and position advance during intervals of time where the fields are constant, it is still important to relate this to the leapfrog time indices in a standard PIC code. 
In most PIC codes, the position and momentum (proper velocity) are staggered in time such that $\mathbf x$ are known at half-integer values of time and $\mathbf u$ are known at integer values of time. 
For a given time step $n$, the fields are assumed constant during the particle push for the interval of time between $n\Delta t$ and $(n+1)\Delta t$; the field values are assumed to be given at time $(n+1/2)\Delta t$, requiring that particle positions are also assumed to be known at time $(n+1/2)\Delta t$. This implicitly assumes that the particle's position does not change during a time step. Under these conditions, we look to analytically advance the momentum forward from time $n\Delta t$ to $(n+1)\Delta t$. Although we may wish to then advance the particle position analytically to time $(n+3/2)\Delta t$ (assuming $d\mathbf x/dt=\mathbf u/\gamma$), this can only be done during time intervals for which $\mathbf u$ is known---only until $(n+1)\Delta t$ for this example. Therefore, the proposed analytical pusher for a standard PIC code is really only doing an analytical advance of the particle momentum. However, the pusher may still lead to significant improvements in accuracy since the momentum advance can lead to much larger errors than the position advance. This is easy to see by noting that the particle's speed is limited by the speed of light, from which it follows that during a time step a particle can only move a fraction of a cell for any field strength. On the other hand, for ultra-strong fields the change in the proper velocity during a time step can be many orders of magnitude, i.e., $\Delta u/u \gg1$, during a time step. 
Nevertheless, the leapfrog advance in the position still leads to noticeable errors compared to an analytic advance in position, as will be shown in a later section.
The use of the analytic solutions together with the pseudo-spectral analytic time domain (PSATD) field solver \cite{vay2013} --- or other concepts \cite{chen2020,higuera2017,lapenta2011} where the position and momentum are defined at the same time --- may lead to new PIC time-indexing algorithms. 

The remainder of the paper is organized as follows: In Section~\ref{sec:soln_eq_motion} and in \ref{sec:eigensys}, we derive the equations for an analytic push of the Lorentz force and introduce the mathematical formalism that can be extended to include the LL and BMT equations. In Section~\ref{sec:soln_LL}, we use the mathematical formalism from \ref{sec:eigensys} to obtain an analytic particle pusher for the 6D phase space including the LL equation. These solutions are exact if the fields are constant during an interval of proper time. In both sections, we also show how to obtain a mapping between the time step in the lab frame and the proper time step. In Section~\ref{sec:soln_BMT}, we derive the analytic solutions to the BMT equation by employing the analytical solutions of momentum obtained in Section~\ref{sec:soln_LL}. The workflow and implementation of the proposed pusher for the 9D phase space are described in Section~\ref{sec:algorithm}. In Section~\ref{sec:sample}, we first show simulation results using the proposed pusher for a single particle in an ultra-intense laser field propagating in vacuum, along with a comparison of results using the standard Boris and Higuera-Cary pushers along with a Boris-like scheme for the spin push. It is shown that the conventional numerical methods lead to large errors in the advance of 9D phase space, while the proposed method provides accurate results. We then conduct full PIC simulations using \textsc{Osiris}~\cite{fonseca2002,hemker2015} to investigate the difference in collective particle behavior using the proposed and conventional pushers. The performance of the proposed and regular pushers is compared in Section~\ref{sec:performance}. A summary and directions for future work are given in Section~\ref{sec:summary}. 

\section{Particle motion in constant and uniform fields without radiation reaction}
\label{sec:soln_eq_motion}

In this section, we will present a derivation of exact solutions to both the momentum and position updates for constant fields.  Analytic expressions can be obtained in various ways. P\'etri~\cite{petri2020} introduced a Lorentz-boosted frame where the $\mathbf E$ and $\mathbf B$ fields are parallel, for which analytic solutions are possible.  The analytic solutions then need to be transformed back to the lab frame. He also provided a mapping between the boosted (proper) and lab frame time steps. Gordon et al.~\cite{gordon2017} showed that the momentum update can be solved analytically in a covariant form and described a matrix representation of the analytic solution. However, Gordon et al. neither provided a mapping between the proper and lab frame time steps nor addressed special cases that need to be considered.  As  noted above, Gordon and Hafizi, ~\cite{gordon2021} very recently proposed a special unitary pusher which provides a second order accurate mapping in the absence of RR. Although the underlying mathematics for obtaining solutions is different, each of the above approaches yields the same net result for the cases considered. However, the forms for the solutions can have different degrees of algorithmic complexity.  Here, we will present another method for finding an analytic expression that is more compact and easier to implement into a PIC code. We use the covariant form for the equations of motion. More importantly, the mathematical formalism we use will be extended to include the LL and the BMT equations in later sections.

The covariant form of the equation of motion without radiation reaction is
\begin{equation}
  \DD{u^\mu}{\tau} = \frac{q}{mc} F^\mu_{~\nu} u^\nu,
\end{equation}
where $u^\mu$ is the four-velocity, $\tau$ is the proper time, $q$ is the particle charge and $m$ is the particle mass. The field tensor is written as
\begin{equation}
\label{eq:field_tensor}
  F^{\mu}_{~\nu}=
  \begin{pmatrix}
    0 & E_1 & E_2 & E_3 \\
    E_1 & 0 & B_3 & -B_2 \\
    E_2 & -B_3 & 0 & B_1 \\
    E_3 & B_2 & -B_1 & 0
  \end{pmatrix}.
\end{equation}
To avoid rewriting constant factors, we use normalized physical quantities, i.e., $\tau\rightarrow\omega_0\tau$, $q\rightarrow \frac{q}{e}$, $m\rightarrow \frac{m}{m_e}$ and $F^\mu_{~\nu}\rightarrow \frac{eF^\mu_{~\nu}}{m_e\omega_0c}$, where $e$ is the elementary charge, $m_e$ is the rest mass of electron and $\omega_0$ is a characteristic reference frequency which, for instance, can be chosen to be the electron plasma frequency or the laser frequency. In addition to the above normalization, we also absorb the charge-to-mass ratio into $F^\mu_{~\nu}$, i.e., $F^\mu_{~\nu}\rightarrow \frac{q}{m}F^\mu_{~\nu}$, to further simplify the expressions. Unless otherwise specified, for the remainder of the paper we will use $F$ to denote the tensor $F^{\mu}_{~\nu}$. The normalized equation of motion is then given by
\begin{equation}
\label{eq:eq_motion}
  \DD{u}{\tau}=Fu.
\end{equation}

If the elements of $F$ are all constant in $\tau$, then it is clear that this equation is easily solved if we know the eigenvalues ($\lambda$) and eigenvectors of $F$. In \ref{sec:eigensys}, it is shown that the field tensor has four eigenvalues that come in pairs. One pair is real, given by $\lambda= \pm\kappa$, and the other pair is purely imaginary, given by $\lambda =\pm\ii\omega$, where
\begin{equation}
\label{eq:eigenval}
  \kappa = \frac{1}{\sqrt{2}}\sqrt{\mathcal{I}_1+\sqrt{\mathcal{I}_1^2+4\mathcal{I}_2^2}},\quad
  \omega = \frac{1}{\sqrt{2}}\sqrt{-\mathcal{I}_1+\sqrt{\mathcal{I}_1^2+4\mathcal{I}_2^2}},
\end{equation}
and
\begin{equation}
\label{eq:invariant}
  \mathcal{I}_1 = |\mathbf E|^2-|\mathbf B|^2,\quad
  \mathcal{I}_2 = \mathbf{E}\cdot\mathbf{B}
\end{equation}
are Lorentz invariants.

In order to obtain general solutions, it is important to project the initial values of the position, $x^\nu$, and proper velocity, $u^\nu$, four-vectors onto the eigenvectors. To facilitate this,  the vector space of $F$ can be split into two subspaces that are each expanded by two eigenvectors, i.e., $\mathbb{S}_\kappa=\text{span}\{e_\kappa, e_{-\kappa}\}$ and $\mathbb{S}_\omega=\text{span}\{e_{\ii\omega}, e_{-\ii\omega}\}$, where $e_\lambda$ is the eigenvector associated with the eigenvalue $\lambda$. It can be shown that (see \ref{sec:eigensys}) $\mathbb{S}_\kappa$ and $\mathbb{S}_\omega$ are mutually orthogonal in the sense of the four-vector inner product. In this article, the four-vector inner product denoted by $(\cdot|\cdot)$ is defined as the contraction of two four-vectors, i.e., $(U|V)=U_\mu V^\mu$ or $(U|V)=U^\text{T}GV$ in the matrix form, where $G\equiv\text{diag}\{1,-1,-1,-1\}$ is the metric tensor. The modulus or length of a four-vector $V$ is thus defined as $|V|\equiv\sqrt{(V|V)}$. In this article, we will decompose some physical quantities into $\mathbb{S}_\kappa$ and $\mathbb{S}_\omega$ to simplify the mathematical derivation. The decomposition or projection can be achieved by applying the projection operator to a physical four-vector of interest, $U$, i.e., $U_\kappa=P_\kappa U$ and $U_\omega=P_\omega U$ where $P_\kappa$ and $P_\omega$ are the projection operators defined as (see \ref{sec:eigensys} and ref. \cite{yaremko2013}),
\begin{equation}
\label{eq:proj_op}
  P_\kappa = \frac{\omega^2I+F^2}{\kappa^2+\omega^2}, \quad
  P_\omega = \frac{\kappa^2I-F^2}{\kappa^2+\omega^2},
\end{equation}
where $I$ is the $4\times4$ identity tensor.

With these definitions, we next explore the evolution of $u_\kappa$ and $u_\omega$ separately. Taking the proper time derivative of both sides of Eq.~(\ref{eq:eq_motion}) and using the properties $F^2u_\kappa=\kappa^2 u_\kappa$ and $F^2u_\omega=-\omega^2 u_\omega$ (see \ref{sec:eigensys}), the equation of motion can be decomposed into
\begin{equation}
\DD{^2 u_\kappa}{\tau^2} = \kappa^2 u_\kappa
\end{equation}
and
\begin{equation}
\DD{^2 u_\omega}{\tau^2} = -\omega^2 u_\omega.
\end{equation}
The solutions are given by
\begin{align}
\label{eq:uk}
  u_\kappa(\tau) &= u_{\kappa0}\cosh(\kappa\tau) + Fu_{\kappa0}\sinc(\ii\kappa\tau)\tau, \\
\label{eq:uo}
  u_\omega(\tau) &= u_{\omega0}\cos(\omega\tau) + Fu_{\omega0}\sinc(\omega\tau)\tau
\end{align}
where $u_{\kappa0}=u_\kappa(\tau=0)$ and $u_{\omega0}=u_\omega(\tau=0)$, with each obtained via $u_{\kappa0}=P_\kappa u_0$ and $u_{\omega0}=P_\omega u_0$. The function $\sinc(x)$ is defined as $\sinc(x)\equiv\sin(x)/x$ and $\sinc(\ii x)\equiv\sin(\ii x)/\ii x=\sinh(x)/x$ is also real-valued.

The four-position can then be obtained by directly integrating the expressions for the proper velocity over the proper time to give
\begin{align}
\label{eq:xk}
x_\kappa(\tau)-x_{\kappa0} &= \left[ u_{\kappa0}\sinc(\ii\kappa\tau) + \frac{1}{2} Fu_{\kappa0}\sinc^2\left(\frac{\ii\kappa\tau}{2}\right) \tau \right]\tau, \\
\label{eq:xo}
x_\omega(\tau)-x_{\omega0} &= \left[ u_{\omega0}\sinc(\omega\tau) + \frac{1}{2} Fu_{\omega0} \sinc^2\left(\frac{\omega\tau}{2}\right) \tau\right]\tau,
\end{align}
where the initial position components $x_{\kappa0}$ and $x_{\omega0}$ are likewise obtained by $x_{\kappa0}=P_\kappa x_0$ and $x_{\omega0}=P_\omega x_0$. 

These equations represent analytic expressions for how to advance the particle four-velocity and four-position from initial to final values during an interval of the proper time in absence of radiation reaction.  The 6D phase space evolution could therefore be advanced during an interval of the proper time, i.e., a proper  time step $\tau$. However, in a PIC simulation, the fields are advanced using the lab-frame time step $\Delta t$. We therefore need to advance forward the phase space for fixed lab-frame steps rather than a fixed proper time step for each particle. Although  $\tau$ changes for each simulation (observer) time step, a mapping between the lab and proper time intervals for fixed fields can be found using the time-like component of Eqs.~(\ref{eq:xk}) and (\ref{eq:xo}):
\begin{equation}
\label{eq:dt_mapping}
\Delta t = x^0(\tau)-x_0^0(\tau) = x^0_{\kappa}(\tau) + x^0_{\omega}(\tau) - x^0_{\kappa0} - x^0_{\omega0},
\end{equation}
or which can be explicitly written as
\begin{equation}
\label{eq:dt_mapping_exp}
  \Delta t = \left( u^0_{\kappa0} \sinc(\ii\kappa\tau) + u^0_{\omega0}\sinc(\omega\tau) 
  + \frac{\mathbf{u}\cdot\mathbf{E}}{2} \left[ \sinc^2\left(\frac{\ii\kappa\tau}{2}\right)
  + \sinc^2\left(\frac{\omega\tau}{2}\right) \right] \tau \right) \tau.
\end{equation}
This is a transcendental equation consisting of trigonometric and hyperbolic functions. We usually need to resort to some root-finding algorithms such as the Newton-Raphson method with second-order precision or the Householder method with higher precision to seek the solution.
We emphasize that the projection operators and Eq.~(\ref{eq:dt_mapping_exp}) are valid as long as $\kappa$ and $\omega$ do not simultaneously vanish. However, Eq.~(\ref{eq:dt_mapping}) is still correct as $\omega$ and $\kappa\rightarrow0$. This special case $\kappa\rightarrow0$ and $\omega\rightarrow0$ will be discussed separately.
In the situation where only $\kappa$ or only $\omega$ vanishes, which indicates $\mathbf{E}$ and $\mathbf{B}$ are mutually orthogonal ($\mathcal{I}_2=0$) but not equal in the amplitude ($\mathcal{I}_1\neq0$), there may be difficulties in evaluating $\sinc(x)$ and $\sinhc(x)$ on a computer due to the singularities. To avoid these practical issues one can Taylor expand them to the machine precision when $x$ is smaller than a specified threshold value $\epsilon_\text{th}$. The technical details of the numerical implementation and performance optimization of the trigonometric and hyperbolic functions can be found in \ref{sect:singularity_app}.

The only special case that needs to be treated separately is the limit that $\mathbf E$ and $\mathbf B$ are mutually orthogonal and equal in magnitude, i.e., $\mathcal{I}_1\rightarrow0$ and $\mathcal{I}_2\rightarrow0$. In this case, both $\kappa$ and $\omega$ approach zero (both are smaller than $\epsilon_\text{th}$) and the subspace decomposition is thus no longer valid. 
To obtain the solution in this case, one can first view $\kappa$ and $\omega$ as small non-zero quantities so that Eqs.~(\ref{eq:uk}) and (\ref{eq:uo}) are still valid. Upon summing Eqs.~(\ref{eq:uk}) and (\ref{eq:uo}), applying the relations $u_{\kappa0}=P_\kappa u_0$ and $u_{\omega0}=P_\omega u_0$, and then Taylor expanding the trigonometric and hyperbolic functions, we are left with
\begin{equation}
\label{eq:u}
  u(\tau) = u_0 + Fu_0\tau + \frac{1}{2}F^2u_0\tau^2 + \frac{1}{6}F^3u_0\tau^3 + O(F^4\tau^4).
\end{equation}
In the above derivation, we have utilized the relations $P_\kappa+P_\omega=I$, $\kappa^2P_\kappa-\omega^2P_\omega = F^2$ and $\kappa^4P_\kappa+\omega^4P_\omega=F^4$. These can be verified using Eqs. (\ref{eq:proj_op}) and (\ref{eq:charcs_eq_F}). Note that Eq. (\ref{eq:u}) can be recast as $u(\tau)=\exp(F\tau)u_0$ which is valid in general. This form for $u(\tau)$ was also mentioned in Gordon et al. \cite{gordon2017}. While this form is not amenable for an algorithm, it is useful for finding $u(\tau)$ in the small $\omega$ and $\kappa$ limit.

The four-position can be similarly obtained by Taylor expanding the sum of Eqs.~(\ref{eq:xk}) and (\ref{eq:xo}),
\begin{equation}
\label{eq:x}
  x(\tau)-x_0 = u_0\tau + \frac{1}{2} Fu_0 \tau^2 + \frac{1}{6} F^2u_0 \tau^3 + \frac{1}{24}F^3u_0\tau^4 + O(F^4\tau^5).
\end{equation}
If we start from $u(\tau)=\exp(F\tau)u_0$ it is trivial to obtain $x(\tau)-x_0=(\exp(F\tau)-I)F^{-1}u_0$ which is the same as Eq. (\ref{eq:x}).
It should be pointed out that $F^n\rightarrow0~(n\ge3)$ when $\kappa,\omega\rightarrow0$ (the proof can be found in \ref{sec:eigensys}), so the first three terms of RHS in Eqs. (\ref{eq:u}) and (\ref{eq:x}) are the exact solutions when $\kappa,\omega\rightarrow0$. In the implementation, we keep terms to order $F^3u_0$ for $\omega,\kappa<\epsilon_\text{th}$.

\section{Particle motion in constant and uniform fields with radiation reaction}
\label{sec:soln_LL}

In this section, we will derive the exact solutions to the LL equation by utilizing the orthogonality of $\mathbb{S}_\kappa$ and $\mathbb{S}_\omega$ introduced in the previous section. We will see that by splitting the four-velocity $u$ into components belonging to the two subspaces, i.e., $u_\kappa$ and $u_\omega$, the integration of the LL equation is greatly simplified.
This subspace decomposition method was also used by Yaremko to obtain the analytical solution of momentum to the LL equation for non-vanishing eigenvalues \cite{yaremko2013}. In this section, we will also discuss the semi-analytical solution of the particle position and proper velocity for the cases with vanishing eigenvalues. The covariant form for the  LL equation can be written as
\begin{equation}
  \DD{u^\mu}{\tau} = \frac{q}{mc}F^\mu_{~\nu}u^\nu + \frac{2q^3}{3m^2c^3}\left(
  \PP{F^\mu_{~\nu}}{x^i}u^\nu u^i
  - \frac{q}{mc^2}F^\mu_{~\nu}F_i^{~\nu}u^i
  + \frac{q}{mc^2}(F^i_{~j}u^j)(F_i^{~k}u_k)u^\mu
  \right),
\end{equation}
where $x^i$ is the four-position. We are investigating cases where the fields are assumed constant in the proper time during a time step. Furthermore, it has been shown by others~\cite{tamburini2010} that the first term in the parentheses with the partial derivatives of $x^i$ can be neglected. This is referred to as the reduced Landau-Lifshitz model. After normalizing all quantities as described in Section~\ref{sec:soln_eq_motion}, the reduced LL equation can be written as
\begin{equation}
\label{eq:reduced_LL}
  \DD{u}{\tau} = Fu + \sigma_0\frac{q^2}{m}\left[ F^2u - (u|F^2 u)u \right],
\end{equation}
where $\sigma_0$ is a dimensionless parameter defined as $\sigma_0=\frac{2e^2\omega_0}{3m_ec^3}$. 

By utilizing the subspace decomposition for $u$, i.e., $u = u_\kappa + u_\omega$, and recalling the relations $F^2 u_\kappa=\kappa^2 u_\kappa$ and $F^2 u_\omega=-\omega^2 u_\omega$, it can be shown that the contraction $(u|F^2u)$ becomes $(u|F^2u)=\kappa^2(u|u_\kappa)-\omega^2(u|u_\omega)=\kappa^2 |u_\kappa|^2 - \omega^2 |u_\omega|^2$. Substituting this result into the reduced LL equation~(\ref{eq:reduced_LL}) and using the fact that the four-velocity has unit length, i.e., $|u|^2=|u_\kappa|^2+|u_\omega|^2=1$, we obtain two decoupled nonlinear differential equations,
\begin{align}
  \label{eq:ode_kappa}
  \DD{u_\kappa}{\tau} &= F u_\kappa + \alpha_0( 1 - |u_\kappa|^2 ) u_\kappa, \\
  \label{eq:ode_omega}
  \DD{u_\omega}{\tau} &= F u_\omega - \alpha_0( 1 - |u_\omega|^2 ) u_\omega,
\end{align}
where $\alpha_0\equiv\sigma_0\frac{q^2}{m}(\kappa^2+\omega^2)$. To solve the nonlinear ordinary differential equation~(\ref{eq:ode_kappa}) [Eq.~(\ref{eq:ode_omega}) can be solved in an analogous manner], we first construct a trial solution as the product of the amplitude of $u_{\kappa}$ and a four-vector $w_\kappa$, i.e., $u_\kappa=|u_\kappa(\tau)| w_\kappa$. This implies that $w_\kappa$ is also enforced to have unit length. With this assumption Eq.~(\ref{eq:ode_kappa}) can be separated into two ODEs as
\begin{align}
\label{eq:pha_kappa}
\DD{w_\kappa}{\tau} &=F w_\kappa, \\
\label{eq:amp_kappa}
\DD{|u_\kappa|}{\tau} &=\alpha_0(1-|u_\kappa|^2)|u_\kappa|.
\end{align}
The first ODE is exactly the unperturbed Lorentz equation. It implies that the modulus of $w_\kappa$ does not change, which can be justified by left multiplying $w_\kappa$ on both sides of the equation and using the property $(w_\kappa|Fw_\kappa)=0$  as described in \ref{sec:eigensys}. As discussed in Section~\ref{sec:soln_eq_motion}, the unperturbed Lorentz equation~(\ref{eq:pha_kappa}) has the solution
\begin{equation}
  w_\kappa(\tau) = w_{\kappa0}\cosh(\kappa\tau) + F w_{\kappa0}\sinc(\ii\kappa\tau) \tau,
\end{equation}
where $w_{\kappa0}=w_\kappa(\tau=0)$. 

Eq.~(\ref{eq:amp_kappa}) can be directly integrated to obtain a  solution to the amplitude equation,
\begin{equation}
  |u_\kappa(\tau)|=\frac{|u_{\kappa0}|}{ \sqrt{|u_{\kappa0}|^2+|u_{\omega0}|^2 e^{-2\alpha_0\tau}} }.
\end{equation}
Combining the solutions for $w_\kappa$ and $|u_\kappa|$ yields
\begin{equation}
\label{eq:uk_rr}
  u_\kappa(\tau) = \frac{1}{ \sqrt{|u_{\kappa0}|^2+|u_{\omega0}|^2 e^{-2\alpha_0\tau}} }
  \left[ u_{\kappa0}\cosh(\kappa\tau) + F u_{\kappa0}\sinc(\ii\kappa\tau) \tau \right].
\end{equation}
The solution to $u_\omega$ can be obtained in an analogous way and is given by
\begin{equation}
\label{eq:uo_rr}
  u_\omega(\tau) = \frac{1}{ \sqrt{|u_{\omega0}|^2+|u_{\kappa0}|^2 e^{2\alpha_0\tau}} }
  \left[ u_{\omega0}\cos(\omega\tau) + F u_{\omega0}\sinc(\omega\tau) \tau \right].
\end{equation}

There is no simple and closed-form expression for the four-position if radiation reaction is included. In reference \cite{yaremko2013} an exact solution was written as an infinite series but it is not amenable to a computational algorithm. However, it is still possible to obtain approximate expressions with sufficiently high accuracy as long as the ``friction'' coefficient $\alpha_0$ is much less than $\tau$. Simple estimates can show that this premise is often true for problems of interest. According to its definition, we know that $\alpha_0=(\frac{4\pi}{3}\frac{r_e}{\lambda_0})\frac{q^4}{m^3}\sqrt{\mathcal{I}_1^2+4\mathcal{I}_2^2} \le (\frac{4\pi}{3}\frac{r_e}{\lambda_0})\frac{q^4}{m^3}(|\mathbf E|^2+|\mathbf B|^2)$, where $r_e$ is the classical electron radius. The equality is true if and only if $\mathbf{E}$ and $\mathbf{B}$ are parallel, i.e., $\mathbf{E}\cdot\mathbf{B}=|\mathbf E||\mathbf B|$. For example, assuming the characteristic length $\lambda_0\sim1~\mu\text{m}$ and the normalized field strengths $E$ and $B$ are on the order of $10^3$, we get $\alpha_0\sim10^{-2}$. In simulations, the time step must be properly selected to sufficiently resolve the characteristic time scales, say $\Delta t\sim0.1$, and thus $\tau\sim\Delta t/\gamma\le1$. Therefore the upper limit of $\alpha_0\tau$ is on the order of $10^{-3}$ when $0<\tau<\Delta \tau$, and keeping only the first term in the  Taylor expansions of the denominator in Eqs.~(\ref{eq:uk_rr}) and (\ref{eq:uo_rr}) is consequently valid. The four-position $x_\kappa$ can be  approximately given by integrating the lowest-order expansion of Eq.~(\ref{eq:uk_rr}),
\begin{equation}
\label{eq:xk_rr}
\begin{aligned}
  x_\kappa(\tau)-x_{\kappa0}
  &= u_{\kappa0}\sinc(\ii\kappa\tau)(1+\alpha_0|u_{\omega0}|^2\tau)\tau
  + \frac{1}{2}(Fu_{\kappa0}-\alpha_0|u_{\omega0}|^2 u_{\kappa0})\sinc^2\left(\frac{\ii\kappa\tau}{2}\right)\tau^2 \\
  &- \alpha_0|u_{\omega0}|^2 Fu_{\kappa0}\Theta(\ii\kappa\tau)\tau^3 + O(\alpha_0^2),
\end{aligned}
\end{equation}
where $\Theta(x)\equiv[\cos(x)-\sinc(x)]/x^2$. Calculating $\Theta(x)$ for $x<\epsilon_\text{th}$ will be discussed in \ref{sect:singularity_app}. It should be noted that $\Theta(\ii x)$ is also a real-valued function of $x$. Similarly, we have
\begin{equation}
\label{eq:xo_rr}
\begin{aligned}
  x_\omega(\tau)-x_{\omega0}
  &= u_{\omega0}\sinc(\omega\tau)(1-\alpha_0|u_{\kappa0}|^2\tau)\tau
  + \frac{1}{2}(Fu_{\omega0}+\alpha_0|u_{\kappa0}|^2 u_{\omega0})\sinc^2\left(\frac{\omega\tau}{2}\right)\tau^2 \\
  &+ \alpha_0|u_{\kappa0}|^2 Fu_{\omega0}\Theta(\omega\tau)\tau^3 + O(\alpha_0^2),
\end{aligned}
\end{equation}
These expressions can then be used to approximately obtain the time step mapping using Eq.~(\ref{eq:dt_mapping}), and the  fast root-finding algorithms mentioned previously in Section~\ref{sec:soln_eq_motion} are still applicable. When the $\sinc(x)$ and $\Theta(x)$ are Taylor expanded, Eqs. (\ref{eq:uk_rr})-(\ref{eq:xo_rr}) are also valid for the situation where only $\kappa$ or $\omega$ approaches to zero.

As previously discussed, the sub-space decomposition fails in the situation where $\kappa\rightarrow0$ and $\omega\rightarrow0$ simultaneously ($\mathbf{E}$ and $\mathbf{B}$ are orthogonal and equal in magnitude). Similarly as we obtain Eq. (\ref{eq:u}), the solution can be sought by first replacing $u_{\kappa0}$ and $u_{\omega0}$ with $P_\kappa u_0$ and $P_\omega u_0$, respectively, summing Eqs. (\ref{eq:uk_rr}) and (\ref{eq:uo_rr}) and then Taylor expanding in terms of $\omega$ and $\kappa$. Moreover, the moduli $|u_{\kappa0}|^2$ and $|u_{\omega0}|^2$ also needs to be expressed in terms of $u_0$. It can be shown (see \ref{sec:mod_4v}) that
\begin{equation}
\label{eq:mod}
  |u_{\kappa0}|^2 = \frac{\omega^2-|Fu_0|^2}{\kappa^2+\omega^2}, \quad
  |u_{\omega0}|^2 = \frac{\kappa^2+|Fu_0|^2}{\kappa^2+\omega^2}.
\end{equation}
Adding Eqs.~(\ref{eq:uk_rr}) and (\ref{eq:uo_rr}) together yields
\begin{equation}
\label{eq:u_rr1}
  u(\tau) = \frac{e^{\frac{1}{2}\alpha_0\tau} P_\kappa\left[u_0\cosh(\kappa\tau) + Fu_0\sinc(\ii\kappa\tau)\tau \right]
  + e^{-\frac{1}{2}\alpha_0\tau}P_\omega \left[u_0\cos(\omega\tau) + Fu_0\sinc(\omega\tau)\tau \right]}{ \sqrt{|u_{\kappa0}|^2 e^{\alpha_0\tau} + |u_{\omega0}|^2 e^{-\alpha_0\tau}} }.
\end{equation}
In order to keep a simple form, we Taylor expand the numerator and denominator separately, rather than seeking a full expansion. Inserting Eqs.~(\ref{eq:mod}) and (\ref{lemma:proj}) into Eq. (\ref{eq:u_rr1}) and then Taylor expanding gives
\begin{equation}
\label{eq:u_rr2}
\begin{split}
  u(\tau) &\simeq \frac{
    u_0 + (Fu_0+\tilde\sigma_0 F^2u_0)\tau + \frac{1}{2}F^2u_0\tau^2
  }{
    \sqrt{1-2\tilde\sigma_0y_0\tau + (\omega^2-\kappa^2)\tilde\sigma_0\tau }
  } \\
  &\quad + \frac{
    \frac{1}{2}(\omega^2-\kappa^2)(u_0+Fu_0\tau)\tilde\sigma_0\tau 
    + F^3u_0\left(\frac{1}{6}\tau^3+\tilde\sigma_0\tau^2\right)
    + O(F^4\tau^4)
  }{
    \sqrt{1-2\tilde\sigma_0y_0\tau + (\omega^2-\kappa^2)\tilde\sigma_0\tau }
  },
\end{split}
\end{equation}
where $\tilde\sigma_0\equiv\sigma_0\frac{q^2}{m}$ and $y_0\equiv|Fu_0|^2$.

Similarly, we can obtain the expression for $x(\tau)$ by adding Eqs.~(\ref{eq:xk_rr}) and (\ref{eq:xo_rr}) and inserting (\ref{eq:mod}) and (\ref{lemma:proj}), giving
\begin{equation}
\begin{split}
  x(\tau)-x_0 &= u_0\tau + \frac{1}{2}\left(Fu_0 + \tilde\sigma_0y_0 u_0 + \tilde\sigma_0F^2u_0\right)\tau^2
  + \frac{1}{3}\left( \tilde\sigma_0y_0 Fu_0 + \frac{1}{2}F^2 u_0\right) \tau^3
  + \frac{1}{8}\tilde\sigma_0y_0 F^2u_0\tau^4 \\
  &\quad + \frac{1}{3}\left(\tilde\sigma_0 + \frac{1}{8}\tau + \frac{1}{10}\tilde\sigma_0y_0\tau^2\right)F^3u_0\tau^3 + O(F^4\tau^5).
\end{split}
\end{equation}

\section{Spin precession in uniform and constant fields}
\label{sec:soln_BMT}

In this section, we will derive the semi-analytic solutions to the particle four-spin vector in uniform and constant fields by utilizing the analytic expression of the four-velocity in absence of RR. After obtaining analytic solutions for the spin evolution based on the analytic evolution of $u$ without RR, we then include RR as two half-impulse split operators at the beginning and end  of each time step. As in the previous sections, we will first discuss the solutions for the general case, followed by special case where the eigenvalues vanish. 

The spin precession of a single charged particle is described by the BMT equation. According to ref. \cite{bargmann1959}, the covariant form of the BMT equatio, is
\begin{equation}
  \DD{s^\mu}{\tau} = \frac{q}{mc}\left[ \frac{g}{2}F^\mu_{~\nu}s^\nu - \frac{1}{c^2}
  \left(\frac{g}{2}-1\right)(u_i F^i_{~j} s^j)u^\mu \right],
\end{equation}
where $g$ is the Land\'e g-factor and is dimensionless. 

The four-spin $s^\mu$ here is described in the observer frame, and hence its time-like component is nonzero. However, as an intrinsic property, it is more conventional to investigate the spin precession dynamics in the particle rest frame. Therefore, we need to transform $s^\mu$ to the particle rest frame after solving the BMT equation. Using normalized units and absorbing the $\frac{q}{m}$ factor into $F$ as done in the previous two sections, the BMT equation can be written as
\begin{equation}
\label{eq:bmt}
  \DD{s}{\tau} = (1+a)Fs - a(u|Fs)u,
\end{equation}
where $a\equiv\frac{g}{2}-1$ is the anomalous magnetic moment ($a\simeq0.0011614$ for electrons). Equation~(\ref{eq:bmt}) is a set of four coupled linear ODEs for spin with variable coefficients due to the presence of the proper velocity terms. If the analytic solutions for the four-velocity in the presence of RR are used, there is no analytic solution to the four-spin. However, we show next that if the analytic solutions for the four-velocity without RR is used then an analytic solution for the spin can be found.

We first explore the time evolution of the scalar $f\equiv (u|Fs)$ and show that it can be analytically solved even without knowing how $s$ evolves. We define $f(\tau)=(u_\kappa|Fs_\kappa) + (u_\omega|Fs_\omega) \equiv f_\kappa(\tau) + f_\omega(\tau)$ and then split Eq.~(\ref{eq:bmt})  into $\mathbb{S}_\kappa$ and $\mathbb{S_\omega}$ based on the eigenvalues of $F$ as was done for the proper velocity:
\begin{align}
  \label{eq:ode_sk}
  \DD{s_\kappa}{\tau} &= (1+a)Fs_\kappa - a f(\tau)u_\kappa, \\
  \label{eq:ode_so}
  \DD{s_\omega}{\tau} &= (1+a)Fs_\omega - a f(\tau)u_\omega.
\end{align}
Combining Eqs.~(\ref{eq:ode_kappa}) and (\ref{eq:ode_sk}) and using the fact that $F^2s_\kappa=\kappa^2 s_\kappa$ and $(u_\kappa|Fu_\kappa)=0$, it follows that the time derivative of $f_\kappa$ is
\begin{equation}
\label{eq:ode_fk}
  \DD{f_\kappa}{\tau} = a\kappa^2 (u_\kappa|s_\kappa).
\end{equation}
Similarly, the time derivative of $f_\omega$ is 
\begin{equation}
\label{eq:ode_fo}
  \DD{f_\omega}{\tau} = -a\omega^2 (u_\omega|s_\omega).
\end{equation}
The quantity $\mathcal{I}_3\equiv \omega^2f_\kappa-\kappa^2 f_\omega$ is an invariant. This can be readily verified by taking the appropriate linear combination  Eqs.~(\ref{eq:ode_fk}) and (\ref{eq:ode_fo}),
\begin{equation}
  \DD{\mathcal{I}_3}{\tau}=a\omega^2\kappa^2
  \left[(u_\kappa|s_\kappa) + (u_\omega|s_\omega)\right]
  =a\omega^2\kappa^2 (u|s) = 0.
\end{equation}
Here we have also used the fact $(u|s)=0$, which follows from the fact that the time-like component of four-spin in the particle rest frame is zero, i.e.,  according to the Lorentz transformation $s'^0=\gamma s^0 - \mathbf{u}\cdot\mathbf{s}\equiv (u|s)=0$. Taking the time derivative of Eq.~(\ref{eq:ode_fk}) and substituting this in Eqs.~(\ref{eq:ode_kappa}) and (\ref{eq:ode_sk})  gives
\begin{equation}
  \DD{^2f_\kappa}{\tau^2} = a^2\kappa^2 \left( |u_\omega|^2f_\kappa - |u_\kappa|^2 f_\omega \right).
\end{equation}
We can similarly get the second-order ODE for $f_\omega$,
\begin{equation}
  \DD{^2f_\omega}{\tau^2} = -a^2\omega^2 \left( |u_\kappa|^2f_\omega - |u_\omega|^2 f_\kappa \right).
\end{equation}
Adding these two ODEs together and using the relations $|u_\kappa|^2+|u_\omega|^2=1$ and $f=f_\kappa+f_\omega$, we finally arrive at
\begin{equation}
\label{eq:ode_f}
  \DD{^2 f}{\tau^2}=-a^2\Omega^2 f + a^2\mathcal{I}_3,
\end{equation}
where $\Omega^2=\omega^2|u_\kappa|^2-\kappa^2|u_\omega|^2=\omega^2|u_{\kappa0}|^2-\kappa^2|u_{\omega0}|^2$ (note that $|u_\kappa|$ and $|u_\omega|$ are constant without RR). It should be noted that $\Omega^2$ is always positive due to $|u_{\kappa}|^2\ge1$ and $|u_{\omega0}|^2\le0$ (see \ref{sec:mod_4v} for the proof). The solution is
\begin{equation}
\label{eq:f}
  f(\tau)=-\Omega^{-2}(\mathcal I_3-f_0\Omega^2)\cos(a\Omega\tau) 
  + (a\Omega)^{-1}\dot f_0\sin(a\Omega\tau) + \Omega^{-2}\mathcal{I}_3,
\end{equation}
where we have used the initial conditions $f_0 = (u_{\kappa0}|Fs_{\kappa0}) + (u_{\omega0}|Fs_{\omega0})$ and $\dot f_0 = a[\kappa^2(u_{\kappa0}|s_{\kappa0})-\omega^2(u_{\omega0}|s_{\omega0})]$. After obtaining the solution to $f(\tau)$, we insert it back into Eqs.~(\ref{eq:ode_sk}) and (\ref{eq:ode_so}) to solve for $s_\kappa$ and $s_\omega$. Eqs.~(\ref{eq:ode_sk}) and (\ref{eq:ode_so}) can now be treated as inhomogeneous ODEs, and the complete solutions are the sum of the homogeneous ($\bar{s}$) and inhomogeneous solutions ($\tilde{s}$), i.e.,
\begin{equation}
s_\kappa=\bar{s}_\kappa+\tilde{s}_\kappa, \quad s_\omega=\bar{s}_\omega+\tilde{s}_\omega.
\end{equation}
The homogeneous solutions satisfy
\begin{equation}
  \ddot{\bar s}_\kappa = (1+a)^2 \kappa^2 \bar s_\kappa, \quad
  \ddot{\bar s}_\omega = -(1+a)^2 \omega^2 \bar s_\omega.
\end{equation}
We impose the initial conditions $\bar{s}_\lambda(0)=s_{\lambda0}$ and $\dot{\bar{s}}_\lambda(0)=(1+a)Fs_{\lambda0}$ ($\lambda=\kappa,\omega$) to the above ODEs, which implies that the inhomogeneous solutions must satisfy the initial conditions $\tilde{s}_\lambda(0)=0$ and $\dot{\tilde{s}}_\lambda(0)=-af_0u_{\lambda0}$ ($\lambda=\kappa,\omega$). Solving the above homogeneous ODEs yields
\begin{align}
  \label{eq:sk_bar}
  \bar{s}_\kappa(\tau) &= s_{\kappa0} \cosh[(1+a)\kappa\tau] + (1+a)Fs_{\kappa0}\sinc[\ii(1+a)\kappa\tau]\tau, \\
  \label{eq:so_bar}
  \bar{s}_\omega(\tau) &= s_{\omega0} \cos[(1+a)\omega\tau] + (1+a)Fs_{\omega0} \sinc[(1+a)\omega\tau]\tau.
\end{align}

Since the inhomogeneous terms in Eqs.~(\ref{eq:ode_sk}) and (\ref{eq:ode_so}) include $u_\kappa$ and $u_\omega$, we can thus construct the trial solution of $\tilde s_\kappa$ as the linear combination of $u_\kappa$ and $\dot u_\kappa$, and that of $\tilde s_\omega$ as the linear combination of $u_\omega$ and $\dot u_\omega$, i.e.,
\begin{equation}
\label{eq:soln_inhom}
  \tilde s_\kappa = C_\kappa(\tau)u_\kappa + D_\kappa(\tau)\dot u_\kappa, \quad
  \tilde s_\omega = C_\omega(\tau)u_\omega + D_\omega(\tau)\dot u_\omega.
\end{equation}
According to the initial conditions to which $\tilde{s}_\kappa$, $\tilde{s}_\omega$ and their time derivatives must be consistent with, we have that the coefficients in Eq.~(\ref{eq:soln_inhom}) must meet the initial conditions $C_\lambda(0)=0$, $\dot C_\lambda(0)=-af_0$, $D_\lambda(0)=0$ and $\dot D_\lambda(0)=0$ ($\lambda=\kappa,\omega$). A set of first-order ODEs for these coefficients can be found by inserting Eq.~(\ref{eq:soln_inhom}) into (\ref{eq:ode_sk}) and (\ref{eq:ode_so}) and comparing the coefficients of $u_\kappa$, $\dot{u}_\kappa$, $u_\omega$ and $\dot{u}_\omega$. Therein we have used $\ddot{u}_\kappa=\kappa^2u_\kappa$ and $\ddot{u}_\omega=-\omega^2u_\omega$. The detailed process for solving these coefficients is tedious and can be found in \ref{sect:part_soln_app}. Here, we directly list the final results. For the general case, we have
\begin{equation}
\label{eq:CD_k}
\begin{split}
  C_\kappa &= a\tau\left[-f_0 \Xi_1(a\Omega\tau, \ii a\kappa\tau)
  + \dot f_0 \tau \Xi_2(a\Omega\tau, \ii a\kappa\tau)
  + a^2\tau^2\mathcal{I}_3 \Xi_3(a\Omega\tau, \ii a\kappa\tau) \right], \\
  D_\kappa &= a^2\tau^2 \left[ f_0\Xi_2(a\Omega\tau, \ii a\kappa\tau)
  + \dot f_0\tau \Xi_3(a\Omega\tau, \ii a\kappa\tau)
  + a^2\tau^2 \mathcal{I}_3 \Xi_4(a\Omega\tau, \ii a\kappa\tau) \right]
\end{split} 
\end{equation}
and
\begin{equation}
\label{eq:CD_o}
\begin{split}
  C_\omega &= a\tau\left[-f_0 \Xi_1(a\Omega\tau, a\omega\tau)
  + \dot f_0 \tau \Xi_2(a\Omega\tau, a\omega\tau)
  + a^2\tau^2\mathcal{I}_3 \Xi_3(a\Omega\tau, a\omega\tau) \right], \\
  D_\omega &= a^2\tau^2 \left[ f_0\Xi_2(a\Omega\tau, a\omega\tau)
  + \dot f_0\tau \Xi_3(a\Omega\tau, a\omega\tau)
  + a^2\tau^2 \mathcal{I}_3 \Xi_4(a\Omega\tau, a\omega\tau) \right]
\end{split}
\end{equation}
where the binary functions $\Xi_i(x,y)~(i=1,...,4)$ are defined as
\begin{equation}
\begin{split}
\Xi_1(x,y) = \frac{x\sin(x)-y\sin(y)}{x^2-y^2}, \quad
&\Xi_2(x,y) = \frac{\cos(x)-\cos(y)}{x^2-y^2}, \\
\Xi_3(x,y) = \frac{\sinc(x)-\sinc(y)}{x^2-y^2}, \quad
&\Xi_4(x,y) = \frac{\sinc^2(x/2)-\sinc^2(y/2)}{2(x^2-y^2)}
\end{split}
\end{equation}

The singularities appearing in these coefficients and in Eqs.~(\ref{eq:sk_bar}) and (\ref{eq:so_bar}) must be treated with care. Apart from the singularities caused by either $\kappa\rightarrow0$ or $\omega\rightarrow0$, the characteristic frequency $\Omega$ in the denominators of Eqs.~(\ref{eq:CD_k}) and (\ref{eq:CD_o}) will also bring about singularities. The singularities of $\Xi_i(x,y)$ functions will be discussed in detail in \ref{sect:singularity_app}. It should be noted that Eqs. (\ref{eq:sk_bar})-(\ref{eq:CD_o}) are also valid for the cases where either $\omega$ or $\kappa$ vanishes.

Just as for the momentum advance, the only case that needs to be  treated specially is when $\omega$ and $\kappa$ simultaneously vanish. In this case, $\Xi_1\rightarrow1$, $\Xi_2\rightarrow-\frac{1}{2}$, $\Xi_3\rightarrow-\frac{1}{6}$ and $\mathcal{I}_3$ vanishes. Therefore, the coefficients are given by
\begin{equation}
  C_\kappa = C_\omega = C = -a\tau\left( f_0 +\frac{1}{2}\dot f_0\tau \right), \quad
  D_\kappa = D_\omega = D = -a^2\tau^2\left( \frac{1}{2}f_0 +\frac{1}{6}\dot f_0\tau \right).
\end{equation}
As previously stated, the subspace decomposition fails in this situation. The homogeneous solution $\bar s$ can be found by adding Eqs.~(\ref{eq:sk_bar}) and (\ref{eq:so_bar}) together, applying the relations $s_{\kappa0}=P_\kappa s_0$ and $s_{\omega0}=P_\omega s_0$ and then taking the limit $\kappa,\ \omega\rightarrow0$, i.e.,
\begin{equation}
\label{eq:s_bar}
  \bar s = s_0 + (1+a)\tau Fs_0 + \frac{1}{2}(1+a)^2\tau^2 F^2s_0 .
\end{equation}
Note that since $F^3=0$ when both $\kappa$ and $\omega$ vanish, Eq. (\ref{eq:s_bar}) actually gives the exact solution.
The inhomogeneous solution $\tilde s$ can be also obtained by simply adding $\tilde{s}_\kappa$ and $\tilde{s}_\omega$ and then evaluating at $\kappa,\ \omega\rightarrow0$, i.e., $\tilde s=C(\tau)u(\tau)+D(\tau)\dot u(\tau)$, where $u(\tau)$ and $\dot u(\tau)$ can be found from Eq.~(\ref{eq:u}).

\section{Algorithm workflow}
\label{sec:algorithm}

\begin{figure}[htbp]
\centering
\includegraphics[width=0.7\textwidth]{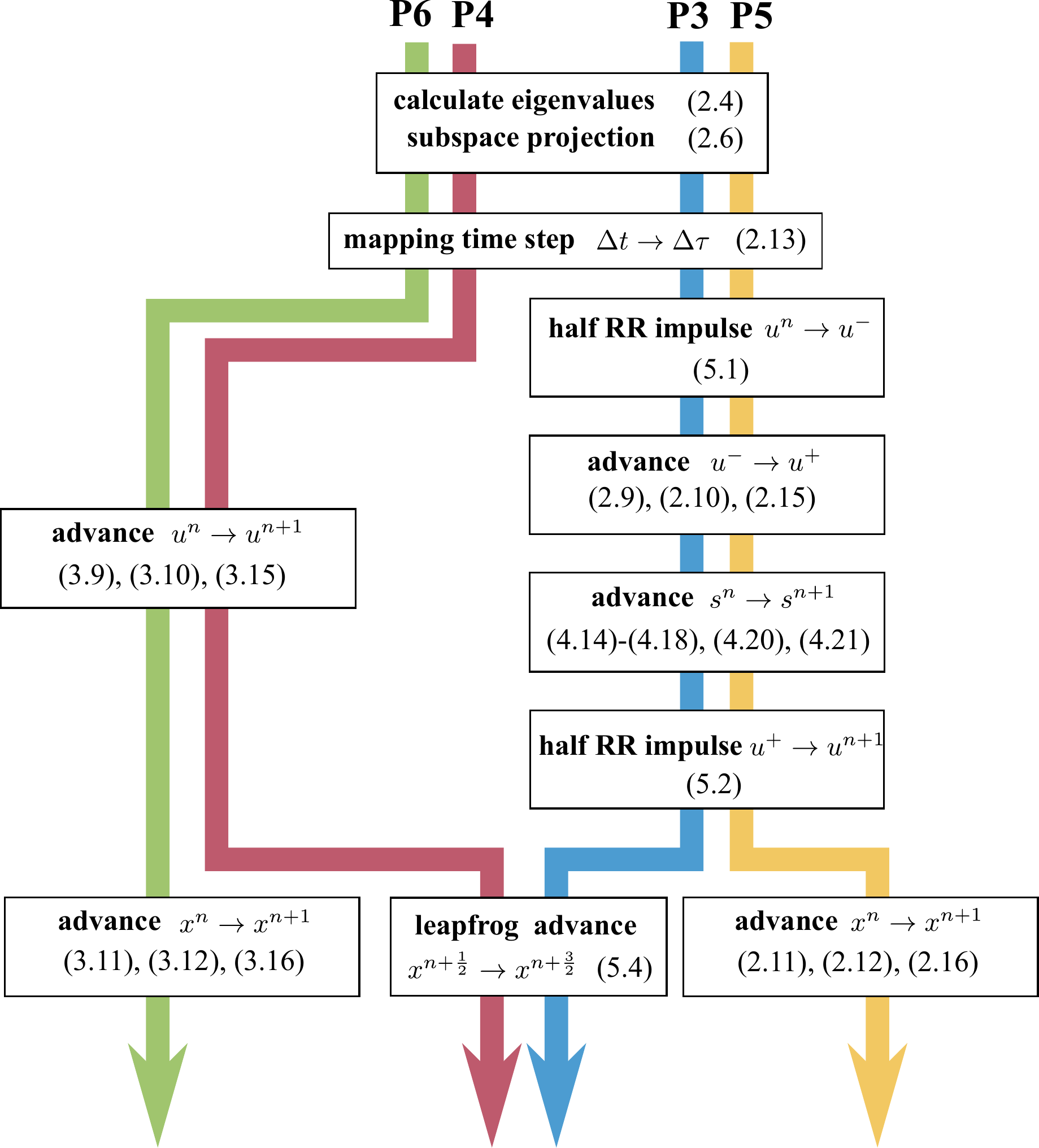}
\caption{Numerical workflow of four algorithm implementations. The relevant equation numbers are summarized in each block, and the blue, red, yellow and green paths correspond to the pusher combinations P3--P6, respectively. The red and green paths analytically advance particle momentum with radiation reaction (RR) included, but without considering spin.  The blue and yellow paths analytically advance particle momentum and spin without considering RR, but incorporate RR by applying two half-impulses at the beginning and end of the advance.  The blue and red paths advance the position through the standard leapfrog scheme used in most PIC codes, whereas the yellow and green paths advance the position analytically (requires a new time-indexing PIC algorithm).}
\label{fig:workflow}
\end{figure}

In this section, we will introduce the algorithm workflow using the analytical expressions of 9D phase space obtained in previous sections. Depending on the problem to be studied, we provide four distinct algorithms that are characterized by different choices from the subset of the analytical solutions. Figure~\ref{fig:workflow} shows the numerical workflow of the four algorithm implementations. The numbers of the requisite equations  for each algorithm have been summarized in each block.

For existing PIC codes, the momentum and position of the particles are staggered in time and the fields are needed at the same time as the position. As a result, only the red and blue paths in Fig.~\ref{fig:workflow} are possible without significantly reworking the PIC algorithm. Maintaining the leapfrog advance of the position thus permits modification of only the momentum update, and the field solve and current deposit do not have to be modified. If updating the spin is unimportant, then the red path is desirable, which uses the analytic solution for the momentum with RR included and the leapfrog advance for the position. The blue path should be used when including spin dynamics, and the details for this method are similar to those of the yellow path described below.

For PIC codes that define position and momentum at the same points in time, the yellow path should be used when the evolution of the full 9D phase space is important; otherwise the green path should be selected.
The yellow path utilizes the analytical solutions to $(\mathbf x, \mathbf u, \mathbf s)$ without RR, but the effects of RR are incorporated  by splitting the change in momentum due to RR into two half-impulses that are applied before and after the analytic solution without RR is used. In the time interval $n\Delta t<t<(n+1)\Delta t$, the first half-impulse can be applied to $\mathbf u$ via
\begin{equation}
  \mathbf u^- = \mathbf u^n + \frac{\Delta t}{2}\mathbf f_\text{RR}(\mathbf u^n),
\end{equation}
after which $\mathbf u^-$ is pushed to $\mathbf u^+$ using the analytical solutions for a full time step. Note that we do not solve for $u^0$ from Eqs. (\ref{eq:eq_motion}) or (\ref{eq:reduced_LL}), rather we use the space component $\mathbf{u}$ to find $u^0$, i.e. $u^0=\sqrt{1+\mathbf{u}^2}$. The quantities $\mathbf x$ and $\mathbf s$ are also analytically advanced a full time step (for the blue path only $\mathbf s$ is analytically advanced), where $\mathbf u^-$ is used for the $u_0$ values in the pertinent equations. The other RR half-impulse is then applied via
\begin{equation}
  \mathbf u^{n+1} = \mathbf u^+ + \frac{\Delta t}{2}\mathbf f_\text{RR}(\mathbf u^+).
\end{equation}
Here the RR force is evaluated as follows:
\begin{equation}
  \mathbf f_\text{RR}(\mathbf u) = \sigma_0\frac{q^2}{\gamma m}[F^2 u-(u|F^2u)u]_\text{spatial},
\end{equation}
where the subscript ``spatial'' refers to the space-like component of a four-vector.

If the position and momentum are staggered in time, the analytical expressions of $\mathbf x$ can no longer be used; in the time interval $n\Delta t<t<(n+1)\Delta t$ where the solution to $\mathbf u$ is known, $\mathbf x$ is known only within $n\Delta t<t<(n+1/2)\Delta t$. Therefore, the positions need to be advanced in the conventional leapfrog manner, i.e.
\begin{equation}
  \mathbf x^{n+\frac{3}{2}} = \mathbf x^{n+\frac{1}{2}} + \mathbf u^{n+1} \Delta t/\gamma^{n+1}.
\end{equation}
The two algorithm implementations shown by the red and blue paths in Fig.~\ref{fig:workflow} use the leapfrog method to update the position. We point out that $\mathbf s$ can be defined on the same time grid points as $\mathbf u$ so that the analytical solutions still applies.

\section{Example simulations}
\label{sec:sample}

In this section, we will compare different particle pushers through a series of particle-tracking simulations where (1) a single particle interacts with an ultra-intense laser pulse in prescribed fields and (2) many particles collectively interact with self-consistent fields in an \textsc{Osiris} PIC simulation. As we have multiple options to advance the particle position, momentum and spin, the following schemes (P1--P6) will be investigated to see how accurately they advance the $(\mathbf x, \mathbf u, \mathbf s)$ phase space:

\begin{enumerate}
\item {\bf P1} -- The Boris pusher is used to advance the particle momentum, and the position is advanced in a leapfrog manner with second-order accuracy in $\Delta t$. The RR force is added according to the splitting method addressed in Section~\ref{sec:algorithm}. Vieira's scheme~\cite{vieira2011} is used to advance the spin.
\item {\bf P2} -- The setup is identical to P1 except the Higuera-Cary pusher~\cite{higuera2017} is used to advance the particle momentum.
\item {\bf P3} (analytical momentum and spin, leapfrog position, impulse RR), {\bf P4} (analytical momentum with RR, leapfrog position, no spin), {\bf P5} (analytical momentum, position and spin, impulse RR) and {\bf P6} (analytical momentum and position with RR, no spin) are the blue, red, yellow and green paths in Fig.~\ref{fig:workflow}, respectively.
\end{enumerate}

\subsection{Single-particle motion in ultra-intense laser fields}

In this section, we compare the various pushers using a particle-tracking code in which the fields are prescribed. This permits using the analytic position update as well. We first consider a one-dimensional case in which a laser pulse propagates in vacuum. Test particles are initialized in front of the laser pulse. The plane-wave laser is linearly polarized in the $\hat 2$-direction and moves in the $\hat 1$-direction. The normalized vector potential is given by
\begin{equation}
\label{eq:analytic_field}
  \mathbf A =  a_0 \cos^2\left(\frac{\pi\phi}{2\omega_0\tau_\text{FWHM}}\right) \cos\phi\ \hat{\mathbf e}_2
\end{equation}
when the phase $\phi\equiv \omega_0 t-k_0x_1$ is within $[-\omega_0\tau_\text{FWHM},~\omega_0\tau_\text{FWHM}]$, and vanishes otherwise. Here, $\tau_\text{FWHM}$ is defined as the full-width-at-half-maximum of the field envelope, $\omega_0$ is the laser frequency and $a_0$ is the strength parameter which is connected with the peak intensity via $a_0=0.86\sqrt{I_0 [10^{18}\text{W/cm}^2]}\lambda_0[\mu\text{m}]$. In all of the following comparisons, a pulse duration of $\tau_\text{FWHM}=50~\omega_0^{-1}$ is chosen, and the field is expressed analytically according to Eq.~(\ref{eq:analytic_field}). We assume the laser wavelength to be 0.8 $\mu\text{m}$ and set the reference frequency to be the laser frequency $\omega_0$ so that the dimensionless radiative damping parameter $\sigma_0\approx 1.474\times10^{-8}$. 

\begin{figure}[htbp]
\centering
\includegraphics[width=\textwidth]{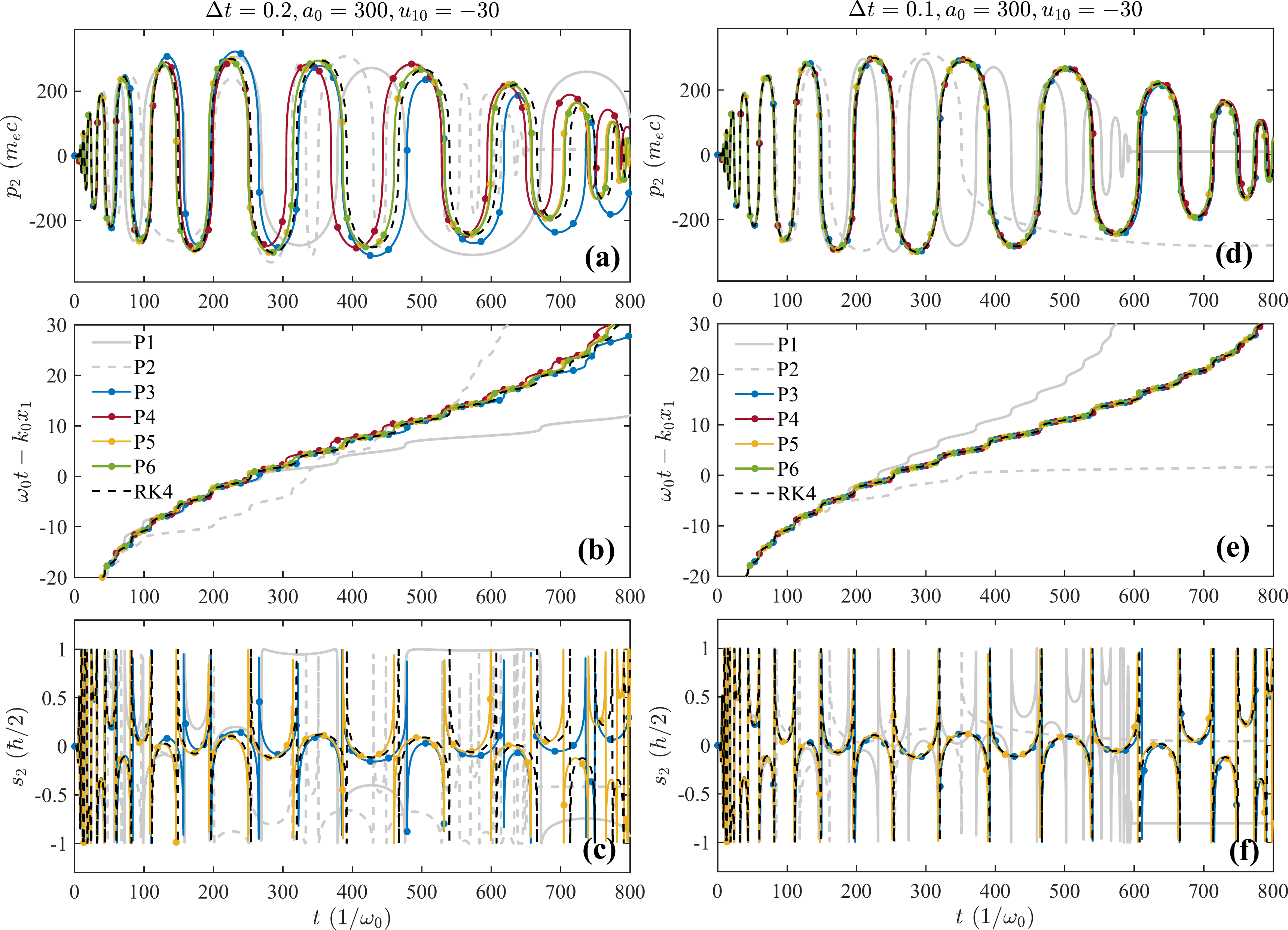}
\caption{Single-particle motion in a head-on collision with an ultra-intense laser ($a_0=300$) using various numerical schemes. The test particle has an initial longitudinal momentum $p_{10}=-30m_ec$. Evolution of (a)(d) transverse momentum $p_2$, (b)(e) phase in laser field and (c)(f) transverse spin $s_2$ are compared for two values of time step. The field felt by the test particle is determined analytically. All proposed pushers give good agreement for $\Delta t = 0.1 \omega_0^{-1}$, whereas P5 and P6 give the best agreement for $\Delta t = 0.2 \omega_0^{-1}$.}
\label{fig:a0_300_u0_m30}
\end{figure}

In the first set of simulations, the test particle has an initial momentum of $p_{10}=-30\ m_ec$ (the negative sign means it counter-propagates relative to the laser), and the initial spin is along the positive $\hat 1$-direction. We tracked the transverse momentum $p_2$, phase $\phi$ and transverse spin $s_2$ during the particle-wave interaction for various values of $\Delta t$. For relatively weak laser intensities where $a_0$ is on the order of unity, it is found that all the aforementioned numerical schemes provide nearly identical and correct phase space trajectories. This is not the case for higher intensities.  
Figure~\ref{fig:a0_300_u0_m30} shows the results for $a_0=300$ ($I_0\sim 2.2\times10^{23}$ W/cm$^2$) for two values of $\Delta t$. The black dashed line is obtained using a fourth-order Runge-Kutta integrator with sufficiently small time step that it can be viewed as the ``correct'' result. It can be seen that the schemes which use the split operator, i.e., standard particle pushers (P1 and P2), lead to incorrect results for both $\Delta t = 0.2\omega_0^{-1}$ and $\Delta t = 0.1\omega_0^{-1}$. The phase shift of particles pushed by P1 and P2 are severely miscalculated [see Figs.~\ref{fig:a0_300_u0_m30}(b) and (e)], which leads to a large deviation in the phase space trajectories. According to our tests, P1 and P2 do not converge until reducing $\Delta t$ to $\sim0.02\omega_0^{-1}$. For P3 and P4, which advance the position in a leapfrog manner and the momentum with the analytical pusher (P3 also analytically advances spin), the momentum and spin oscillations and phase shift are qualitatively correct, but quantitatively inaccurate for $\Delta t=0.2\omega_0^{-1}$ [see Figs.~\ref{fig:a0_300_u0_m30}(a)--(c)]. When the time step is reduced to $\Delta t=0.1\omega_0^{-1}$, both P3 and P4 converge to the ``correct'' results as shown in Figs.~\ref{fig:a0_300_u0_m30}(d)--(f). Since P5 and P6 advance both position and momentum analytically (P5 also advances spin analytically), they give good agreement with the ``correct'' results for the two time steps, as expected.

\begin{figure}[htbp]
\centering
\includegraphics[width=\textwidth]{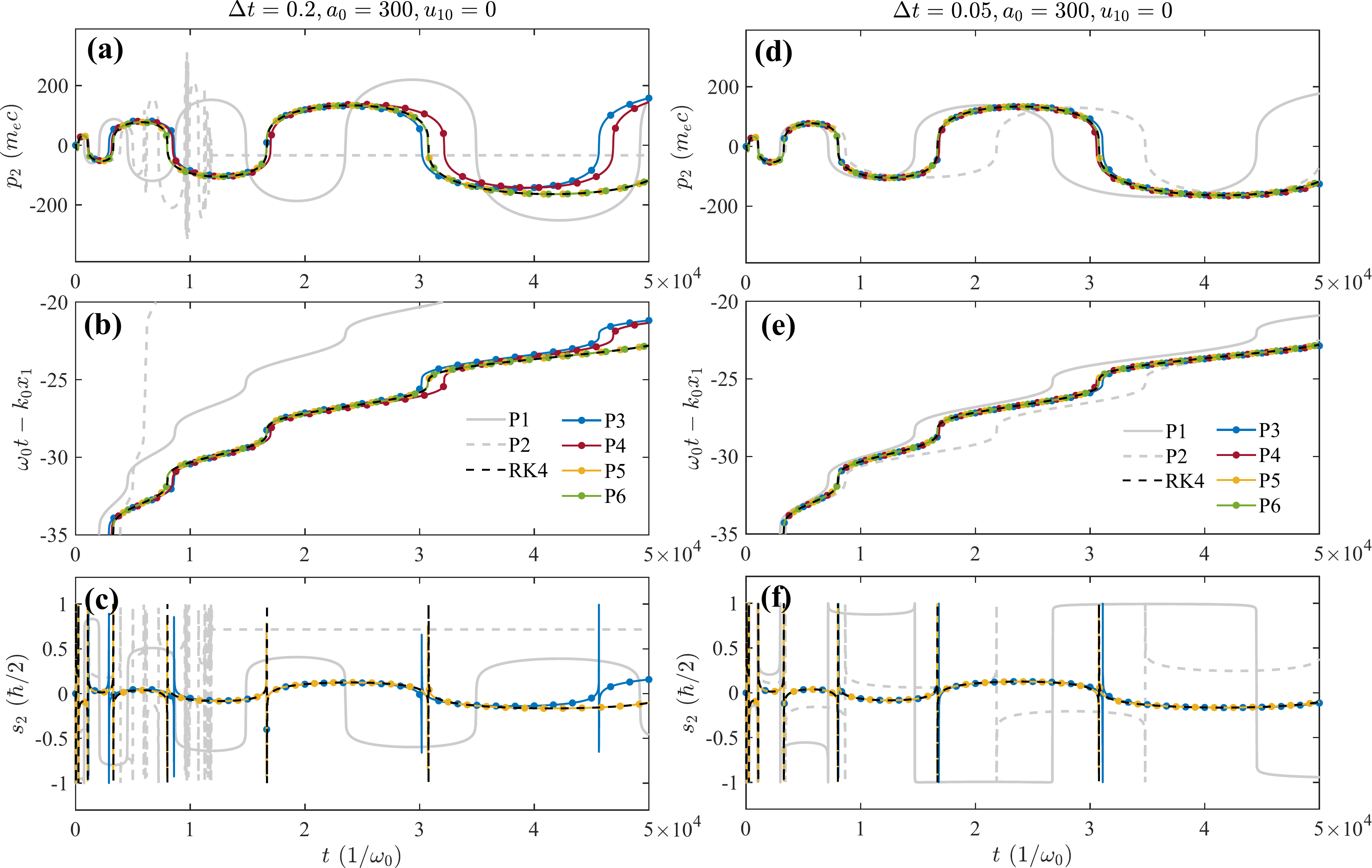}
\caption{Single-particle motion in an ultra-intense laser ($a_0=300$) using various numerical schemes. The test particle is initialized at rest. Evolution of (a)(d) transverse momentum $p_2$, (b)(e) phase in laser field and (c)(f) transverse spin $s_2$ are compared for two values of time step. The field felt by the test particle is determined analytically. All proposed pushers give good agreement for $\Delta t = 0.05 \omega_0^{-1}$, whereas P5 and P6 give the best agreement for $\Delta t = 0.2 \omega_0^{-1}$.}
\label{fig:a0_300_u0_0}
\end{figure}

We next tested how well these numerical schemes work with zero initial momentum as shown in Fig.~\ref{fig:a0_300_u0_0}. According to Vranic et al.~\cite{vranic2016b}, this situation is more sensitive to numerical noise. Due to the energy loss during the laser-particle interaction, the particle will stay in phase for much longer, increasing the duration of interaction. We tested two time steps, $\Delta t=0.2\omega_0^{-1}$ and $\Delta t=0.05\omega_0^{-1}$, and again P1 and P2 lead to a large deviation from the correct results. For $\Delta t=0.2\omega_0^{-1}$, P3 and P4 lead to  the correct phase space trajectory results for the first few cycles, but clear deviations appear at  later times due to the accumulation of  numerical errors over a long duration. Good agreement can be reached when the time step is reduced to $\Delta t=0.05\omega_0^{-1}$. As before, P5 and P6 lead to excellent agreement with the correct results even for a time step typically used to accurately solve for the fields in laser-plasma-interaction simulations ($\Delta t=0.2\omega_0^{-1}$).

\begin{figure}[htbp]
\centering
\includegraphics[width=\textwidth]{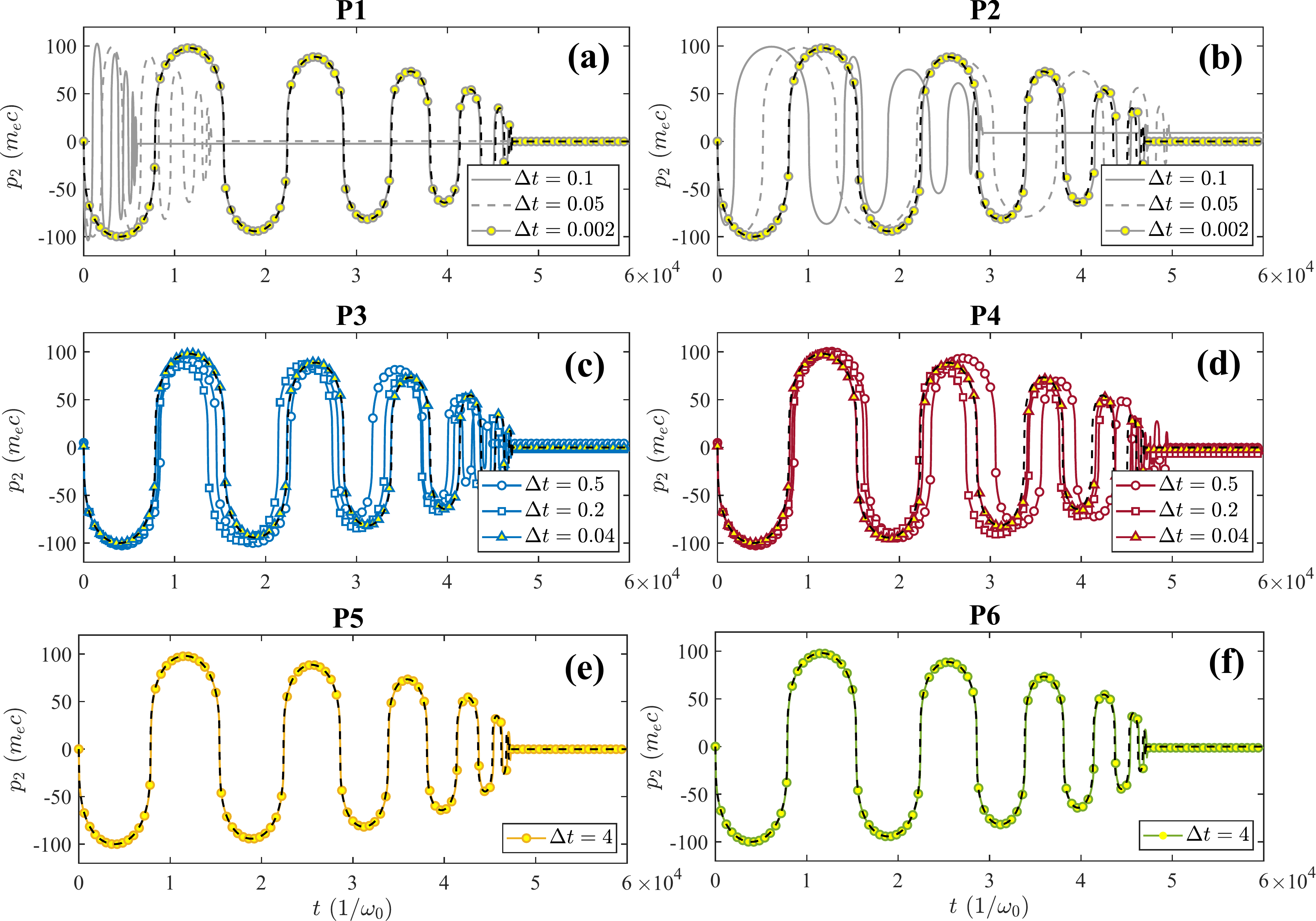}
\caption{Evolution of transverse momentum $p_2$ in a plane-wave ultra-intense laser ($a_0=100$) using various numerical schemes and values of time step. The  particle is initialized at rest where the vector potential of the laser is zero. The black dashed line represents the result of using fourth-order Runge-Kutta method with very high precision, and the time step is reduced for each scheme until convergence is reached.}
\label{fig:conv_test}
\end{figure}

We also examined the effect of using the proposed schemes for the situation where the test particles are initialized inside the laser field. A stationary ($\bm p_0=0$) test particle was initialized inside laser fields with $a_0=100$ at a location where the laser electric field (vector potential) reaches a maximum (zero), i.e., $\phi_0=0$. The evolution of the transverse momentum $p_2$ is shown in Fig.~\ref{fig:conv_test}. We gradually reduced the time step of each scheme to examine the maximum $\Delta t$ for which the simulation result converges to that of the high-precision Runge-Kutta method (black dashed lines in Fig.~\ref{fig:conv_test}). As we can see from Figs.~\ref{fig:conv_test}(a) and (b), the schemes using the regular pushers do not converge at a conventionally selected $\Delta t$ ($\Delta t=0.1$ and $0.05 \omega_0^{-1}$) that resolves the laser frequency, but require an extremely small time step $\Delta t=0.002\omega_0^{-1}$ to converge. The maximum time step for P3 and P4 to converge is around $\Delta t=0.04\omega_0^{-1}$, as shown in Figs.~\ref{fig:conv_test}(c) and (d). Therefore, the benefit of solely using the analytic momentum advance is twenty-fold compared to P1 and P2. However, P5 and P6 converge at an even larger time step $\Delta t=4\omega_0^{-1}$ for this specific problem, as shown in Figs.~\ref{fig:conv_test}(e) and (f), giving a hundred-fold improvement over P3 and P4. It should be noted that the comparison here is to show that the proposed pushers can greatly reduce the requirement for time steps, rather than to give a rule of thumb for choosing a time step. The choice of time step and the benefits of using the proposed pushers are problem-specific.

\subsection{Full PIC simulation of beam-laser interactions}

As shown in the last section, single-particle motion in strong laser fields varies significantly when different particle pushers are used, even when prescribed (analytical) fields are used for the laser. In this section, we will show that the collective behavior of a particle bunch can also vary significantly depending on the choice of the pusher unless very small time steps are used.   
We have implemented the proposed particle pusher into \textsc{Osiris}. The aforementioned P3 is adopted because it uses a time-staggering layout for the particle positions and momenta, along with the resulting need for a leapfrog advance of the particle position.
In the full 2D PIC simulations, a 0.8-$\mu\text{m}$ wavelength bi-Gaussian laser pulse with $a_0=500$, 30 $c/\omega_0$ focal spot size and 50 $\omega_0^{-1}$ FWHM duration for the field envelope collides head-on with an electron beam that travels at an incident angle of 15 degrees. The electron beam has a bi-Gaussian density distribution with rms transverse size $\sigma_\perp=10k_0^{-1}$, rms longitudinal size $\sigma_\parallel=15k_0^{-1}$ and initial momentum $p_{\parallel0}=10m_ec$. The beam has zero emittance and energy spread. The cell sizes are $\Delta x_1=0.2k_0^{-1}$ and $\Delta x_2=2k_0^{-1}$, and the time step is $\Delta t=0.1\omega_0^{-1}$. To accurately simulate the particle motion in the laser field, we have used a Maxwell solver with an extended stencil~\cite{li2021} to reduce the numerical errors arising from numerical dispersion and the interlacing of $E$ and $B$ fields in time.

\begin{figure}[htbp]
\centering
\includegraphics[width=0.7\textwidth]{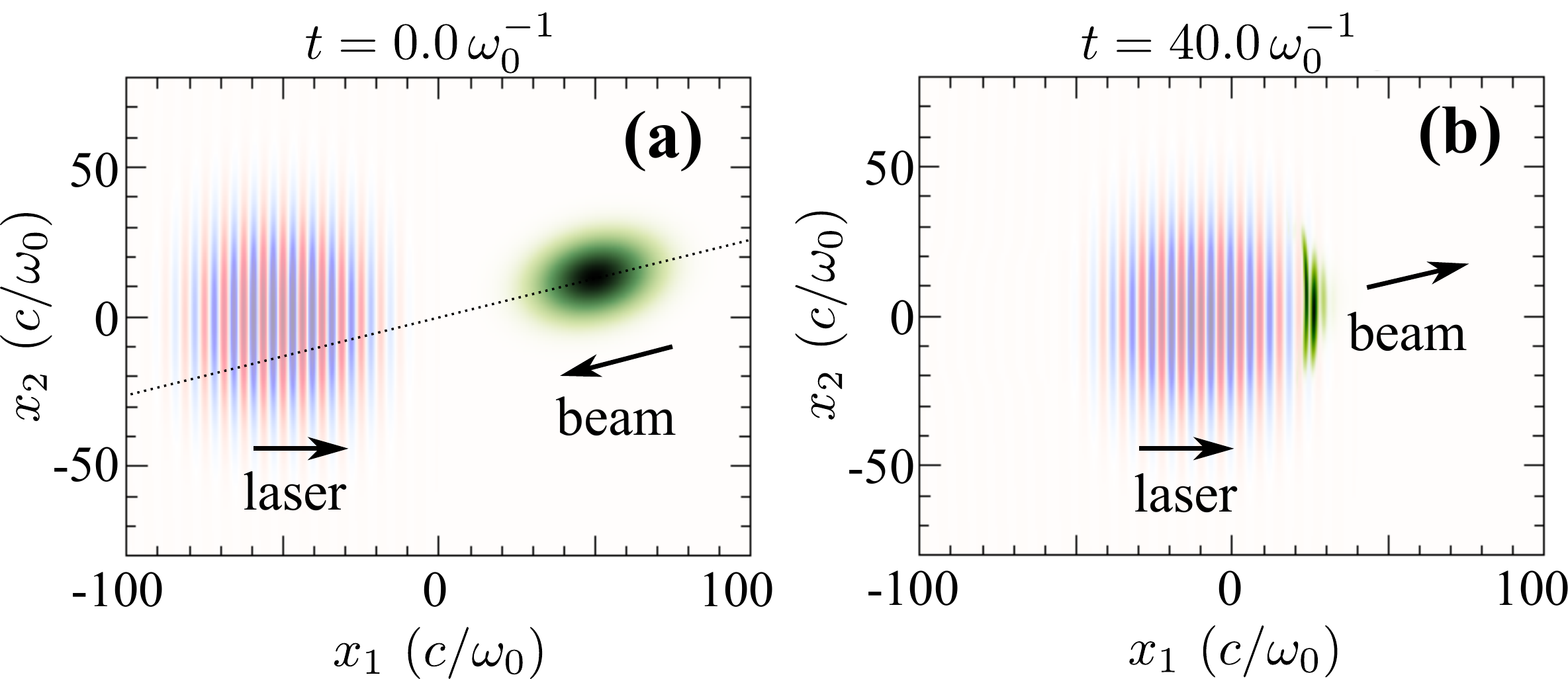}
\caption{2D PIC simulation results using \textsc{Osiris}. Snapshots of the beam density (green) and laser field (red and blue) are shown at (a) $t=0$ and (b) $t=40\omega_0^{-1}$. The bunch length is compressed and reversed by the extremely strong radiation pressure of the laser. The scheme P3 is used to generate the plot.}
\label{fig:collision-sketch}
\end{figure}

\begin{figure}[htbp]
\centering
\includegraphics[width=0.7\textwidth]{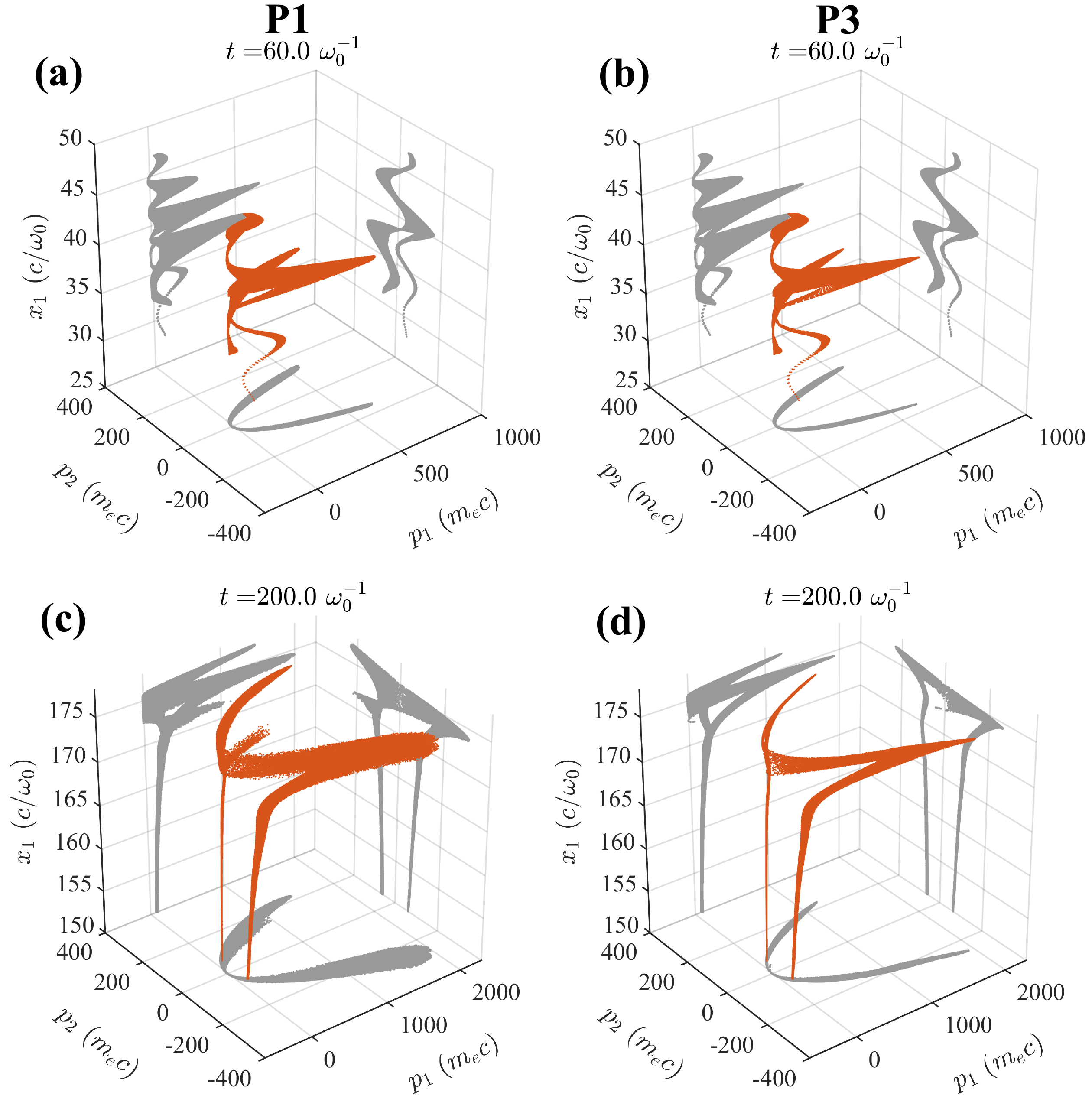}
\caption{Particle distributions in $x_1$-$p_1$-$p_2$ phase space for schemes (a)(c) P1 and (b)(d) P3 at two different times. The phase space distribution remains narrow late in time when using P3, as is expected.}
\label{fig:x1p1p2}
\end{figure}

Figure~\ref{fig:collision-sketch} shows the laser field and beam density distribution. As shown in Fig.~\ref{fig:collision-sketch}(a), the electron beam initially moves toward the laser pulse from right to left. The bunch length is then compressed by the extremely strong radiation pressure of the leading edge of the laser.
The propagation direction of the beam is eventually reversed so that it co-moves with the laser pulse as shown in Fig.~\ref{fig:collision-sketch}(b). There are significant differences in phase space between the ``standard'' and the proposed numerical schemes. Figure~\ref{fig:x1p1p2} shows the $x_1$-$p_1$-$p_2$ space phase for P1 [Figs.~\ref{fig:x1p1p2}(a) and (c)] and P3 [Figs.~\ref{fig:x1p1p2}(b) and (d)] at $t=60\omega_0^{-1}$ and $t=200\omega_0^{-1}$. At $t=60\omega_0^{-1}$ the differences between the two schemes are hardly observable. At $t=200\omega_0^{-1}$ the phase space distribution begins to broaden for P1 while it remains narrow for P3.
We also conducted convergence tests using P1 with ten-fold higher resolution in space and time. The results converged with those shown in Fig.~\ref{fig:x1p1p2}(d) for the larger time step using P3.

\begin{figure}[htbp]
\centering
\includegraphics[width=\textwidth]{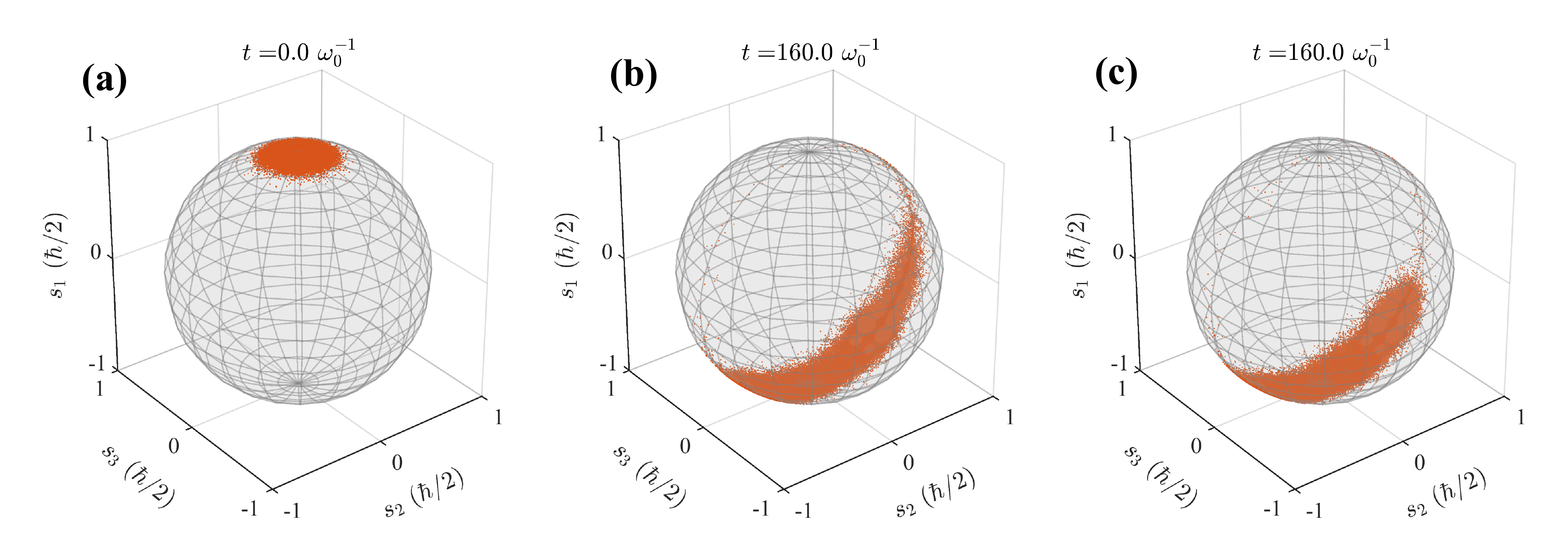}
\caption{Particle distributions in $s_1$-$s_2$-$s_3$ space at (a) $t=0$ and (b)(c) $t=160\omega_0^{-1}$. P1 and P3 were used to obtain the results in (b) and (c), respectively.}
\label{fig:spin_space}
\end{figure}

We also compared how the evolution of the spin precession is modified for the different schemes. The spin of the electron beam is initially polarized along the positive $\hat 1$-direction with a small divergence, as shown in Fig.~\ref{fig:spin_space}(a). In Fig.~\ref{fig:spin_space} we plot the spin in the rest frame so that all the particles move on the surface of a sphere of radius $\hbar/2$ in $s_1$-$s_2$-$s_3$ space. When the beam starts to interact with the laser field, the particles move down toward the negative $\hat 1$-direction along the longitudes. Significant differences between the schemes can be seen at $t=160\omega_0^{-1}$: the particles advanced analytically in momentum space and with the exact spin pusher in spin space using P3 [see Fig.~\ref{fig:spin_space}(c)] are more concentrated at the pole in the negative $\hat 1$-direction, while the particles advanced by the Boris pusher and the Vieira scheme using P1 [see Fig.~\ref{fig:spin_space}(b)] are spread over a much wider region around the pole.

\section{Performance}
\label{sec:performance}

\subsection{Performance optimization}
\label{sec:performance_opt}

Since the calculation of the trigonometric and hyperbolic (T/H) functions is much more computationally expensive than the arithmetic of floating-point operations, the most critical issue for improving the algorithm is to reduce the number of calculations required to evaluate the T/H functions as much as possible. This includes avoiding duplicated computation and utilization of the sum/difference formulas of the T/H functions. The details of the optimization methods can be found in \ref{sec:optimization_app}. In its current implementation, the pushers with the splitting RR correction (P3 and P5) only call the Fortran built-in T/H functions $2l$ times each time step, where $l$ is the number of iterations used in the root-finding procedure of $\tau$. For a relative tolerance of $10^{-3}$ to $10^{-6}$, $l=1$ or $l=2$ is usually sufficient for the iteration to converge to the desired accuracy. For the momentum and position advances, no Fortran built-in T/H function calls are needed. The spin advance of P3 and P5 requires only 3 extra calls of the built-in T/H functions. The analytical pushers with RR (P4 and P6) require $2l$ built-in T/H function calls and $l$ built-in exponential function calls (whose overhead is comparable to that of T/H function). Except for the T/H and exponential function calculations, the number of the remaining floating point arithmetic operations are comparable to that of Boris scheme.

Another performance-related issue is the branching used in our algorithm to deal with the light-like solution ($\kappa\rightarrow0,\omega\rightarrow0$) and the singularity of $\sinc(x)$, $\Theta(x)$ and $\Xi_i(x)$ functions for small arguments. In a PIC simulation there are very few macro-particles that trigger the special branches because it is rare that the electric and magnetic fields felt by almost any particles are ``exactly'' perpendicular to each other and/or exactly equal in magnitude. In this situation, the branch predictor, which is turned on by default by most compilers with -O3 optimization, can work perfectly to prevent the flow in the instruction pipeline from being interrupted. This feature, which is utilized by almost all modern CPU architectures such as x86, can significantly improve the effective performance. According to our tests, there is almost no extra overhead when the branch predictor is on.

\subsection{Performance test}

\begin{figure}[htbp]
\centering
\includegraphics[width=\textwidth]{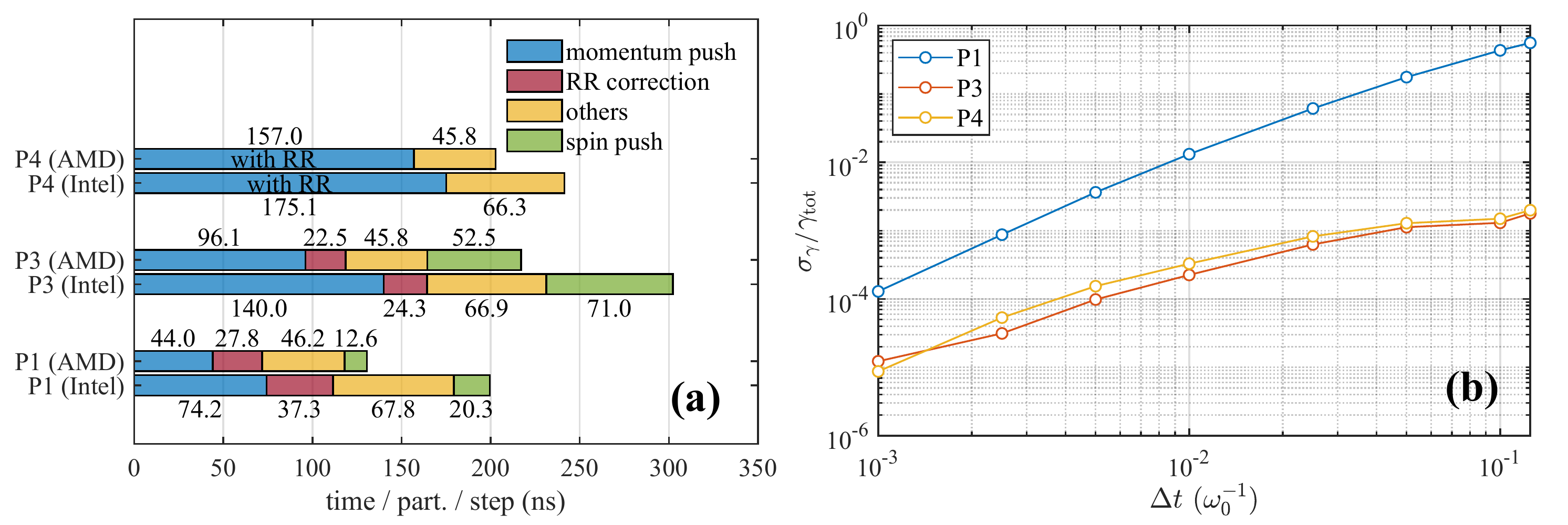}
\caption{(a) One-step performance of various pushers implemented in \textsc{Osiris}. The ``other'' category includes field interpolation, update of particle positions and current deposition. The tests were carried out on Intel Xeon CPU E5-2698 @ 2.3GHz and AMD Ryzen 7 3700X @ 3.6GHz processors. (b) Relative error of the total particle energy vs. the time step for various pushers.}
\label{fig:performance}
\end{figure}

In this section, we compare the performance of the P3 and P4 implementations into \textsc{Osiris} against each other and against the standard Boris push. We carried out two-dimensional simulations of a uniform plasma in ultra-intense standing-wave fields formed by two counter-propagating plane-wave laser along $\hat{1}$-direction. The normalized electromagnetic fields are given by $E_2=B_3=2a_0\sin(\omega_0t)\cos(k_0 x_1)$, where $a_0=500$ is used in this test. The simulation box was $157\times157$ cells large, and the cell size was $0.2k_0^{-1}\times0.2k_0^{-1}$. Each cell contains 4 macro-particles with a particle shape corresponding to quadratic weighting/interpolation. The macro-particles are initialized as $\exp[-u_j^2/(2u_\text{th}^2)]$ where $j=x_1,x_2,x_3$ and $u_\text{th}=5c$ in each direction. \textsc{Osiris} was compiled using GNU Fortran~8.3 with -O3 optimization on an both Intel Xeon E5-2698 and AMD Ryzen 7 3700X processors. For each time step, the time cost of various procedures including the momentum advance, RR correction and spin advance, along with position advances, field interpolation and current deposit (``others'') are summarized in Fig.~\ref{fig:performance}(a). The computational cost of the momentum advance in P3 is 1.9/2.2 times that of P1 (Boris scheme), and the spin advance in P3 is 3.5/4.2 times that of P1 (Vieira scheme) on Intel/AMD platforms. With the RR correction included, the computational cost of the momentum advance of P3 is only 1.5/1.7 times that of P1 on Intel/AMD platforms. The momentum advance, including the RR in P4, is 2.4/3.6 times slower than P1. The additional cost of updating the positions, interpolating fields from the grid onto particle positions and depositing the current onto the grid is shown by the yellow blocks in Fig.~\ref{fig:performance}.

For simulations where particle spin is not considered, the schemes employing exact momentum pushers can provide competitive performance on a per-time-step basis compared to schemes using regular pushers; this includes weak/moderate field scenarios where the regular pushers remain accurate at conventionally selected time steps. However, in moderate- to strong-field regimes, time steps 10--100 times smaller are required for regular pushers to obtain the accuracy of the analytical pushers. We have performed a series of numerical convergence tests to illustrate this point. Fig.~\ref{fig:performance}(b) shows the relative errors of the total particle energy as a function of time step. The relative error is calculated as $\sigma_\gamma/\gamma_\text{tot}=|\gamma_{\text{tot},\Delta t}-\gamma_\text{tot}|/\gamma_\text{tot}$, where $\gamma_{\text{tot},\Delta t}$ is the total particle energy of simulations with time step $\Delta t$ and $\gamma_\text{tot}$ is the reference one with $\Delta t=5\times10^{-4}\omega_0^{-1}$. The relative errors were calculated at $t=30~\omega_0^{-1}$. In the example in this section, for a typical time step $\Delta t=0.1\omega_0^{-1}$ that approaches the Courant limit, the relative errors of P3 and P4 are only $\sim 10^{-3}$ as shown by the red and yellow lines in Fig. \ref{fig:performance}(b), indicating the simulation result is already well converged. However, the P1 scheme needs $\Delta t\sim 3\times10^{-3}\omega_0^{-1}$ to achieve comparable accuracy. Although the momentum push (blue) for P3 and P4 are around 1.6 and 3.0 times (the mid-value of Intel and AMD platforms) slower than P1 for a single time step according the performance test results shown in Fig. \ref{fig:performance}(a), the entire particle loop is only 1.3 and 1.5 times slower. Most importantly the effective speedup is around 31 and 27 times respectively due to much lower requirement for the time step.
Therefore, these new pushers can significantly reduce the computational time needed for high-fidelity simulations in the strong field regime. We note that the performance differences will become even smaller as the order of the particle shape increases, since the steps encapsulated in yellow are the same across schemes and will take longer per particle.

\section{Conclusion}
\label{sec:summary}

In this article, we derived the analytic solutions to the change in the four-vector of momentum and position while including a reduced form of the radiation reaction (RR). We thank the referee for bringing to our attention relevant theoretical work \cite{yaremko2013}. When the equations of motion are written in covariant form, analytic solutions  can be found straightforwardly if the electric and magnetic fields are considered constant (in both space and time) over a single time step. We obtained forms of the solutions to both the momentum (proper velocity) and position [$(\mathbf x,\ \mathbf u)$ phase space] using projection operators amenable to PIC codes. The trajectory of $(\mathbf x,\ \mathbf u)$ can be accurately computed in the strong-field regime with these explicit, closed-form expressions using much larger time steps than would be required for standard pushers. These expressions are analytic, so any errors arise only from the assumption of constant and uniform fields at each time step. When the RR is involved, these expressions are still highly accurate, except for in cases where classical theory fails.

With an analytical solution to $\mathbf u$ and keeping the fields constant and uniform, the Bargmann-Michel-Telegdi equation can also be analytically solved, and the closed-form solutions can be used to simulate spin precession in strong fields. Although these expressions are only perfectly accurate without RR, the effect can still be properly taken into account by separately including radiative impulse corrections to $\mathbf u$. This semi-analytical approach can also be used when RR is modeled as a QED process.

The advantage in computational efficiency (defined as the computational time to accurate solution) of the proposed 9D phase space pusher over existing schemes was demonstrated through a series of single-particle simulations where the fields are associated with a laser. It is shown that the proposed pusher can yield correct or sufficiently accurate phase space trajectories with time steps an order of magnitude smaller than for the standard split operator pushers for normalized laser amplitudes $a_0$ on the order of at least $10^2$.
We note that for problems where the fields vary slowly in time (including high-amplitude imposed magnetic fields), the proposed pusher will be even more efficient than standard schemes.   For example, when the  laser fields are known (given) such that the position can also be analytically updated, the full analytic pusher can in some cases obtain accurate results for $\omega \Delta t=4$ while the standard split operator pushers require $\omega \Delta t=.002$. In this case the particle moved forward with the laser so that it saw a very small Doppler shirted frequency.

We implemented the analytic solution for the momentum update into the code \textsc{Osiris}, maintaining the leapfrog position advance. Therefore, only the momentum update needed to be modified while the field solver, position update and current deposit remained unchanged.  
Using \textsc{Osiris}, PIC simulations were also conducted to compare the proposed numerical scheme against standard schemes for the head-on collision of a spin polarized electron beam with an ultra-intense laser pulse. The results showed significant differences in the phase space (including the spin precession) between the proposed and the standard schemes. As the time step was reduced, the standard pusher simulations converged to that of the analytical pusher case with the larger time steps. Although these sample simulations were all conducted in the context of laser-plasma interactions, the proposed algorithm itself is general and can be applied to many other research fields.

Future work may involve the development of PIC algorithms that define the position and momentum at the same time or that use predictor-corrector algorithms. We found that the proposed scheme without (with) the spin advance is only 20 (80) percent slower per particle than standard pushers (including field interpolation, momentum update, and current deposit) for linear particle shapes.
However, the proposed scheme can provide accurate solutions with time steps much larger than those required for standard pushers (depending on the field strength and configuration), generating significant speedups. For example, for some of the laser-plasma interaction examples presented here where the laser fields are updated using the field solver, time steps as much as 10 times larger can be used with the proposed scheme.

\section*{Acknowledgments}
This work was supported in parts by the US Department of Energy contract number  DE-SC0010064, DE-SC0019010 and SciDAC FNAL subcontract 644405, Lawrence Livermore National Laboratory subcontract B634451, and US National Science Foundation grant numbers 1806046.
The work of MV was supported by the European Research Council (ERC-2015-AdG Grant No. 695088) and Portuguese Science Foundation (FCT) Grant No. SFRH/BPD/119642/2016.
Simulations were carried out on the Cori Cluster of the National Energy Research Scientific Computing Center (NERSC).
We also acknowledge useful comments from the reviewers.

\begin{appendix}
\section{Eigensystem of the field tensor $F$ and relevant properties}
\label{sec:eigensys}

The eigenvalues of the field tensor $F$ (under the assumption that the elements are constant in $\tau$) are determined by the characteristic equation $\text{det}(F-\lambda I)=0$. This leads directly to the following equations for the eigenvalues, 
\begin{equation}
\label{eq:charcs_eq}
\lambda^4 - \mathcal{I}_1 \lambda^2 - \mathcal{I}_2^2 = 0,
\end{equation}
where $\mathcal{I}_1$ and $\mathcal{I}_2$ are the well-known Lorentz invariants \cite{jackson2007},
\begin{equation}
  \mathcal{I}_1 = |\mathbf E|^2-|\mathbf B|^2,\quad
  \mathcal{I}_2 = \mathbf{E}\cdot\mathbf{B}.
\end{equation}
From this it follows that there are two pairs of eigenvalues, $\lambda=\pm\kappa$ and $\lambda=\pm\ii\omega$, where
\begin{equation}
  \kappa = \frac{1}{\sqrt{2}}\sqrt{\mathcal{I}_1+\sqrt{\mathcal{I}_1^2+4\mathcal{I}_2^2}},\quad
  \omega = \frac{1}{\sqrt{2}}\sqrt{-\mathcal{I}_1+\sqrt{\mathcal{I}_1^2+4\mathcal{I}_2^2}}.
\end{equation}

To facilitate the derivations of the analytic pushers, we introduce two subspaces that are defined by the eigenvectors, i.e., $\mathbb{S}_\kappa=\text{span}\{e_\kappa,e_{-\kappa}\}$ and $\mathbb{S}_\omega=\text{span}\{e_{\ii\omega},e_{-\ii\omega}\}$, where $e_\lambda$ denotes the eigenvector associated with the eigenvalue $\lambda$. For general four-vectors $V_\kappa\in\mathbb{S}_\kappa$ and $V_\omega\in\mathbb{S}_\omega$, the following relations,
\begin{equation}
\label{lemma:subspace}
F^2 V_\kappa = \kappa^2 V_\kappa,\quad F^2 V_\omega = -\omega^2 V_\omega
\end{equation}
are satisfied.
These relations can be easliy verified by expressing $V_\kappa$ and $V_\omega$ as a linear combination of the appropriate eigenvectors and then using the fact that $Fe_\lambda=\lambda e_\lambda$. To  decompose an arbitrary four-vector $V$ into $\mathbb{S}_\kappa$ and $\mathbb{S}_\omega$ subspaces, $V$ is rewritten as $V=V_\kappa+V_\omega$, and then the operator $F^2$ is applied to both sides. These  two equations can then be solved for $V_\kappa$ and $V_\omega$ as
\begin{equation}
\label{lemma:proj}
V_\kappa = (\omega^2V + F^2 V) / (\kappa^2+\omega^2),\quad
V_\omega = (\kappa^2V - F^2 V) / (\kappa^2+\omega^2),
\end{equation}
which indicates that $P_\kappa\equiv(\kappa^2+\omega^2)^{-1}(\omega^2I+F^2)$ and $P_\omega\equiv(\kappa^2+\omega^2)^{-1}(\kappa^2I-F^2)$ are the projection operators of a four-vector into $\mathbb{S}_\kappa$ and $\mathbb{S}_\omega$.

According to the Cayley-Hamilton theorem~\cite{householder2013}, the field tensor $F$ also satisfies the characteristic equation~(\ref{eq:charcs_eq}), i.e., $F^4-\mathcal{I}_1 F^2-\mathcal{I}_2^2 I=0$, which leads to
\begin{equation}
\label{eq:charcs_eq_F}
(\kappa^2 I - F^2)(\omega^2 I + F^2) = 0.
\end{equation}
With this property, we can prove that $\mathbb{S}_\kappa$ and $\mathbb{S}_\omega$ are mutually orthogonal  by explicitly taking the inner product of $V_\kappa$ and $V_\omega$ and then substituting in Eq.~(\ref{lemma:proj}) to give
\begin{equation}
(\kappa^2+\omega^2)^2 (V_\kappa|V_\omega) 
= V^\text{T}(\omega^2 I + (F^2)^\text{T} )G(\kappa^2 I - F^2)V 
= V^\text{T}G(\omega^2 I + F^2 )(\kappa^2 I - F^2)V = 0.
\end{equation}
We have also used Eq.~(\ref{eq:charcs_eq_F}) and an obvious relation between the field tensor and its transpose, $F=-GF^\text{T}G$.

Another important relation that will be frequently used in this article is that $F^3=0$ when $\mathcal{I}_1=\mathcal{I}_2=0$ (or $\kappa=\omega=0$). This can be shown by explicit calculation using Eq.~(\ref{eq:field_tensor}) to give
\begin{equation}
  F^3 = \mathcal{I}_1 F + \mathcal{I}_2 F^*,
\end{equation}
where $F^*$ is the dual tensor defined as
\begin{equation}
  F^* =
  \begin{pmatrix}
    0 & B_1 & B_2 & B_3 \\
    B_1 & 0 & -E_3 & E_2 \\
    B_2 & E_3 & 0 & -E_1 \\
    B_3 & -E_2 & E_1 & 0
  \end{pmatrix}.
\end{equation}
Therefore, $F^3$ vanishes when $\mathcal{I}_1=0$ and $\mathcal{I}_2=0$.

\section{Modulus of four-velocity}
\label{sec:mod_4v}

In this appendix, we will discuss the nature of the modulus of the four-velocity components in $\mathbb{S}_\kappa$ and $\mathbb{S}_\omega$. As addressed in \ref{sec:eigensys}, the subspace components $u_\kappa$ and $u_\omega$ can be obtained by projecting $u$ to the subspaces using the projection operators $P_\kappa$ and $P_\omega$, i.e.,
\begin{equation}
  u_\kappa = \frac{\omega^2 u+F^2u}{\kappa^2+\omega^2},\quad
  u_\omega = \frac{\kappa^2 u-F^2u}{\kappa^2+\omega^2}.
\end{equation}
Combining this with the characteristic equation for $F$ [Eq.~(\ref{eq:charcs_eq_F})] written as $F^4=(\kappa^2-\omega^2)F^2+\kappa^2\omega^2I$, the modulus of $u_\kappa$ and $u_\omega$ can be calculated as
\begin{equation}
\label{eq:mod_uk_uo}
  |u_\kappa|^2 = \frac{\omega^2-|Fu|^2}{\kappa^2+\omega^2}, \quad
  |u_\omega|^2 = \frac{\kappa^2+|Fu|^2}{\kappa^2+\omega^2}.
\end{equation}

Now we will prove that the modulus of the four-force has a maximum of $-\kappa^2$.
The problem can be more accurately defined for a given $F$ by finding the extrema of $|Fu|^2$ under the restricted condition $|u|^2=1$. We use the Lagrange multiplier method to handle this problem and construct the Lagrangian function $\mathcal{L}(u,\chi)=|Fu|^2+\chi(|u|^2-1)$, where the scalar $\chi$ is the Lagrange multiplier. The extremum point $(u^*,\chi^*)$ is determined by $\partial_u\mathcal{L}=0$ and $\partial_\chi\mathcal{L}=0$. The latter equation directly gives the restricted condition $|u|^2=1$, and the former can be written in the matrix form as
\begin{equation}
\label{eq:prob_min}
  \frac{1}{2}\PP{\mathcal{L}}{u} = F^\text{T}GFu + \chi Gu = 0.
\end{equation}
The existence of a non-trivial solution for $u$ requires $\text{det}(F^\text{T}GF+\chi G)=0$, from which $\chi^*$ can be determined. Using the fact that $F^\text{T}G=-GF$ and $G$ has a non-zero determinant [$\text{det}(G)=-1$], we have $\text{det}(F^2-\chi I)=0$, which is exactly the characteristic equation of $F^2$; the solution $\chi^*$ is the associated eigenvalue. Recalling that $F$ has two pairs of eigenvalues, $\pm\kappa$ and $\pm\ii\omega$, the characteristic equation therefore has two roots, $\chi^*=\kappa^2$ and $\chi^*=-\omega^2$. Noticing that $(u^*,\chi^*)$ satisfies Eq.~(\ref{eq:prob_min}), the meaning of $\chi^*$ can be revealed by left multiplying Eq.~(\ref{eq:prob_min}) by $u^{*\text{T}}$, which gives $|Fu^*|^2=-\chi^*$. This indicates that $-\kappa^2$ and $\omega^2$ are two extrema of $|Fu|^2$. However, the extremum $\omega^2$ should be discarded because $|Fu|^2<0$ always holds, which can be briefly proved as follows:
\begin{equation}\nonumber
  |Fu|^2 = (\mathbf{u}\cdot\mathbf{E})^2 - |\gamma\mathbf{E}+\mathbf{u}\times\mathbf{B}|^2
  = \dot{\gamma}^2 - |\dot{\mathbf{u}}|^2
  = \frac{(\mathbf{u}\cdot\dot{\mathbf{u}})^2}{\gamma^2} - |\dot{\mathbf{u}}|^2
  \le |\dot{\mathbf{u}}|^2\left( \frac{|\mathbf{u}|^2}{\gamma^2} - 1 \right) < 0.
\end{equation}
Therefore, $|Fu|^2$ has the unique extremum $-\kappa^2$, and we can verify $|Fu|^2\le-\kappa^2$ by substituting in an arbitrary $u$. With this property we can know from Eq.~(\ref{eq:mod_uk_uo}) that
\begin{equation}
  |u_\kappa|^2 \ge 1, \quad |u_\omega|^2 \le 0.
\end{equation}

\section{Inhomogeneous solutions to Eqs.~(\ref{eq:ode_sk}) and (\ref{eq:ode_so})}
\label{sect:part_soln_app}

In this appendix, we will seek the inhomogeneous solutions to Eqs.~(\ref{eq:ode_sk}) and (\ref{eq:ode_so}). Notice that the inhomogeneous terms in Eqs.~(\ref{eq:ode_sk}) and (\ref{eq:ode_so}) contain $u_\kappa$ and $u_\omega$, respectively, so the trial solutions can be constructed  as
\begin{align}
  \label{eq:trial_sk_part}
  \tilde s_\kappa &= C_\kappa(\tau)u_\kappa + D_\kappa(\tau)\dot u_\kappa, \\
  \label{eq:trial_so_part}
  \tilde s_\omega &= C_\omega(\tau)u_\omega + D_\omega(\tau)\dot u_\omega.
\end{align}
Inserting the trial solution of Eq.~(\ref{eq:trial_sk_part}) back into Eq.~(\ref{eq:ode_sk}) and comparing the coefficients of terms proportional to $u_\kappa$ and $\dot u_\kappa$, we get two ODEs for $C_\kappa(\tau)$ and $D_\kappa(\tau)$,
\begin{align}
  \label{eq:ode_CD1}
  \dot C_\kappa(\tau) &= a\kappa^2 D_\kappa(\tau) - af(\tau), \\
  \label{eq:ode_CD2}
  \dot D_\kappa(\tau) &= aC_\kappa(\tau).
\end{align}
Substituting Eq.~(\ref{eq:f}) into above equations, we can find out the solutions that satisfy the zero initial conditions, i.e., $C_\kappa(0)=0$ and $D_\kappa(0)=0$. We note that the initial conditions of $\dot C_\kappa$ and $\dot D_\kappa$ required by the inhomogeneous solutions, i.e., $\dot C_\kappa(0)=-af_0$ and $\dot D_\kappa(0)=0$, are naturally satisfied according to Eqs.~(\ref{eq:ode_CD1}) and (\ref{eq:ode_CD2}). The solutions are given by
\begin{equation}
\begin{split}
  C_\kappa &= \dot f_0\frac{ \cos(a\Omega\tau)-\cosh(a\kappa\tau) }{a(\kappa^2+\Omega^2)}
  + \frac{ h_\Omega\kappa\sin(a\Omega\tau) - h_\kappa\Omega\sinh(a\kappa\tau) }{\kappa\Omega(\kappa^2+\Omega^2)}, \\
  D_\kappa &= \dot f_0\frac{ \kappa\sin(a\Omega\tau)-\Omega\sinh(a\kappa\tau) }{a\kappa\Omega(\kappa^2+\Omega^2)}
  - \frac{ h_\Omega\kappa^2\cos(a\Omega\tau) + h_\kappa\Omega^2\cosh(a\kappa\tau) }{\kappa^2\Omega^2(\kappa^2+\Omega^2)} + \frac{\mathcal{I}_3}{\kappa^2\Omega^2},
\end{split} 
\end{equation}
where $h_\kappa \equiv \mathcal{I}_3 + f_0\kappa^2$ and $h_\Omega \equiv \mathcal{I}_3 - f_0\Omega^2$.

The ODEs of $C_\omega$ and $D_\omega$ can be similarly established by inserting Eq.~(\ref{eq:trial_so_part}) into Eq.~(\ref{eq:ode_so}) and comparing the coefficients:
\begin{align}
  \label{eq:ode_CD3}
  \dot C_\omega(\tau) &= -a\omega^2 D_\omega(\tau) -af(\tau), \\
  \label{eq:ode_CD4}
  \dot D_\omega(\tau) &= aC_\omega(\tau).
\end{align}
The solutions that satisfy $C_\omega(0)=0$ and $D_\omega(0)=0$ are
\begin{equation}
\begin{split}
  C_\omega &= \dot f_0\frac{ \cos(a\omega\tau)-\cos(a\Omega\tau) }{a(\omega^2-\Omega^2)}
  + \frac{ h_\omega\Omega\sin(a\omega\tau) - h_\Omega\omega\sin(a\Omega\tau) }{\omega\Omega(\omega^2-\Omega^2)}, \\
  D_\omega &= \dot f_0\frac{ \Omega\sin(a\omega\tau)-\omega\sin(a\Omega\tau) }{a\omega\Omega(\omega^2-\Omega^2)}
  - \frac{ h_\omega\Omega^2\cos(a\omega\tau) - h_\Omega\omega^2\cos(a\Omega\tau) }{\omega^2\Omega^2(\omega^2-\Omega^2)} - \frac{\mathcal{I}_3}{\omega^2\Omega^2},
\end{split}
\end{equation}
where $h_\omega \equiv \mathcal{I}_3 - f_0\omega^2$.

\section{Algorithm optimization through reducing trigonometric/hyperbolic function calculation}
\label{sec:optimization_app}

As addressed in Sec. \ref{sec:performance_opt}, the part that impacts the performance most is the calculation of the trigonometric and hyperbolic (T/H) functions. Therefore, the key point of the optimization is how to reduce the number of T/H function calculations in a single time step.

First, we should avoid duplicated of T/H function calculations as much as possible. For example, in the root-finding procedure for $\tau$ [Eq. (\ref{eq:dt_mapping_exp})] we need to calculate the terms $\cosh(\kappa\tau)$, $\sinh(\kappa\tau)$, $\cos(\omega\tau)$ and $\sin(\omega\tau)$, where the $\cosh(\kappa\tau)$ and $\cos(\omega\tau)$ are used to calculate $\sinc^2(\ii\kappa\tau/2)$ and $\sinc^2(\omega\tau/2)$ in Eq. (\ref{eq:dt_mapping_exp}) via the identities $2[1-\cos(x)]/x^2=\sinc^2(x/2)$ and $2[\cosh(x)-1]/x^2=\sinc^2(\ii x/2)$. The root-finding subroutine should also output these terms along with the resultant proper time step $\tau$, so that in the momentum and position advance, i.e., Eqs. (\ref{eq:uk})-(\ref{eq:xo}), the T/H function terms will no longer be calculated repeatedly. This optimization technique has also been applied in the analytical pusher with RR (Sec. \ref{sec:soln_LL}).

In the spin advance, we need to first calculate the terms $\cos(a\Omega\tau)$, $\sin(a\Omega\tau)$, $\cos(a\omega\tau)$, $\sin(a\omega\tau)$, $\cosh(a\kappa\tau)$ and $\sinh(a\kappa\tau)$ for the $\Xi_i$ functions to evaluate the coefficients in Eqs. (\ref{eq:CD_k}) and (\ref{eq:CD_o}). Then, these terms can be reused to calculate the $\cosh[(1+a)\kappa\tau]$, $\sinc[\ii(1+a)\kappa\tau)$, $\cos[(1+a)\omega\tau]$ and $\sinc[(1+a)\omega\tau]$ terms of $\bar{s}_\kappa$ and $\bar{s}_\omega$, i.e., Eqs. (\ref{eq:sk_bar}) and (\ref{eq:so_bar}), via the sum and difference formula. Noting that the terms $\cosh(\kappa\tau)$, $\sinh(\kappa\tau)$, $\cos(\omega\tau)$ and $\sin(\omega\tau)$ have already been obtained in the root-finding procedure of $\tau$, thus the calculation of Eqs. (\ref{eq:sk_bar}) and (\ref{eq:so_bar}) does not involve extra direct calls of the T/H functions. When calculating $\tilde{s}_\kappa$ and $\tilde{s}_\omega$, i.e., Eq. (\ref{eq:soln_inhom}), the $u_\kappa$, $\dot{u}_\kappa$, $u_\omega$ and $\dot{u}_\omega$ should use the results already obtained in the momentum advance rather than being recalculated.

Second, as already seen, the cosh/sinh and cos/sin always appear in pairs, thus we can use the following relations
\begin{equation}
\nonumber
  \sinh(x)=2At, \quad \cosh(x)=A(1+t^2), \quad t=\tanh(x/2), \quad A=(1-t^2)^{-1}
\end{equation}
and
\begin{equation}
\nonumber
  \sin(x)=2At, \quad \cos(x)=A(1-t^2), \quad t=\tan(x/2), \quad A=(1+t^2)^{-1}
\end{equation}
to fastly calculate the function pairs. With this technique, the number of T/H function calculation can be further halved.

In summary, by applying the above optimization techniques, we only need to call the Fortran built-in T/H functions to calculate $\tanh(\kappa\tau/2)$ and $\tan(\omega\tau/2)$ once each iteration of the root-finding subroutine of $\tau$. For the analytical pusher with RR, we need to call the Fortran built-in exponential function once in the iteration. In the spin advance, we only need to call the built-in T/H functions three times for the calculation of $\tan(a\Omega\tau/2)$, $\tanh(a\kappa\tau/2)$ and $\tan(a\omega\tau/2)$. No direct call of built-in T/H functions is needed for the remaining elements of the algorithm.

\section{Evaluating functions $\sinc(z)$, $\Theta(z)$ and $\Xi_i(z_1,z_2)$ near singularities}
\label{sect:singularity_app}

To eliminate the singularities of $\sinc(z)$ and $\Theta(z)$ and $\Xi_i(z_1,z_2)~(i=1,...,4)$, we Taylor expand them around the singular point and truncate to the machine precision. In the context of this article, $z$, $z_1$ and $z_2$ are taken as either real or purely imaginary, and it can be verified that $\sinc(z)$, $\Theta(z)$ and $\Xi_i(z_1,z_2)$ are all real-valued according to the definitions.

The Taylor expansion of $\sinc(z)$ and $\Theta(z)$ are
\begin{equation}
\nonumber
\sinc(z) = 1 - \frac{z^2}{6} + \frac{z^4}{120} + O(z^6), \quad
\Theta(z) = -\frac{1}{3} + \frac{z^2}{30} - \frac{z^4}{840} + O(z^6).
\end{equation}

One of the singular points of $\Xi_i(z_1,z_2)$ is $z_1,z_2\rightarrow0$. At this point, $\Xi_i(z_1,z_2)$ can be expanded as
\begin{align}
\nonumber
\Xi_1(z_1,z_2) &=  1 - \frac{1}{6}(z_1^2+z_2^2) + \frac{1}{120}(z_1^4+z_1^2z_2^2+z_2^4) + O(z_1^n z_2^{6-n}), \\
\nonumber
\Xi_2(z_1,z_2) &= -\frac{1}{2} + \frac{1}{24}(z_1^2+z_2^2) - \frac{1}{720}(z_1^4+z_1^2 z_2^2+z_2^4) + O(z_1^n z_2^{6-n}), \\
\nonumber
\Xi_3(z_1,z_2) &= -\frac{1}{6} + \frac{1}{120}(z_1^2+z_2^2) - \frac{1}{5040}(z_1^4+z_1^2 z_2^2+z_2^4) + O(z_1^n z_2^{6-n}), \\
\nonumber
\Xi_4(z_1,z_2) &= -\frac{1}{24} + \frac{1}{720}(z_1^2+z_2^2) - \frac{1}{40320}(z_1^4+z_1^2 z_2^2+z_2^4) + O(z_1^n z_2^{6-n}).
\end{align}
When both $z_1$ and $z_2$ are real numbers, say $z_1=x_1$ and $z_2=x_2$, there is another singular point $x_1\rightarrow x_2\neq0$. We can express $x_1=X+\Delta$ and $x_2=X-\Delta$ where $X=\frac{x_1+x_2}{2}$ and $\Delta=\frac{x_1-x_2}{2}$. Expanding $\Xi_i(x_1,x_2)$ in terms of $\Delta$ yields
\begin{align}
\nonumber
\Xi_1(x_1,x_2) &= \left(\frac{1}{2} - \frac{\Delta^2}{12} \right)\cos(X)
+ \left(\frac{1}{2} - \frac{\Delta^2}{4} \right)\sinc(X) + O(\Delta^4), \\
\nonumber
\Xi_2(x_1,x_2) &= \left(-\frac{1}{2} + \frac{\Delta^2}{12}\right)\sinc(X) + O(\Delta^4), \\
\nonumber
\Xi_3(x_1,x_2) &= \left[ \frac{1}{2X^2} - \frac{(X^2-6)\Delta^2}{12X^4} \right]\cos(X) 
- \left[ \frac{1}{2X^2} - \frac{(X^2-2)\Delta^2}{4X^4} \right]\sinc(X) + O(\Delta^4), \\
\nonumber
\Xi_4(x_1,x_2) &= -\frac{1}{X^4} - \frac{2}{X^6} 
+ \left[ \frac{1}{X^4} - \frac{(X^2-4)\Delta^2}{2X^6} \right]\cos(X)
+ \left[ \frac{1}{2X^2} - \frac{(X^2-18)\Delta^2}{12X^4} \right]\sinc(X) + O(\Delta^4).
\end{align}

\end{appendix}

% \section*{References}
% \begin{thebibliography}{00}
% \end{thebibliography}

\bibliographystyle{elsarticle-num}
\bibliography{refs}

\end{document}